\documentclass[twocolumn]{jpsj3}
\usepackage{txfonts}
\RequirePackage{graphicx,color}

\title{
Theory of Magnetoacoustic Resonance to Probe Multipole Effects \\
Due to a Crystal Field Quartet
}

\author{Mikito Koga$^1$ and Masashige Matsumoto$^2$}

\inst{
$^1$Department of Physics, Faculty of Education, Shizuoka University, Shizuoka
422--8529, Japan \\
$^2$Department of Physics, Faculty of Science, Shizuoka University, Shizuoka
422--8529, Japan \\
}

\inst{
$^1$Department of Physics, Faculty of Education, Shizuoka University, Shizuoka
422--8529, Japan \\
$^2$Department of Physics, Faculty of Science, Shizuoka University, Shizuoka
422--8529, Japan \\
}

\abst{
We present a new method of acoustically driven resonance that probes octupole degrees of
freedom as well as a quadrupole usually hidden by the magnetic properties of a crystal field
quartet.
A characteristic of the quadrupole is reflected in the anisotropic resonance transition rate, which
depends on the propagation direction of a surface acoustic wave under an external magnetic field
parallel to a typical crystallographic axis.
The transition rate is modulated by the anisotropic Zeeman splitting associated with octupoles.
We demonstrate how to obtain information about  the quartet quadrupole--strain coupling and
evaluate the anisotropic octupole effect quantitatively.
We also discuss the applicability of our method to identifying a quadrupole order parameter using
a multipole--multipole interaction model.
For large excitation energy gaps under strong magnetic fields, we propose a photon-assisted magnetoacoustic resonance formulated on the basis of the Floquet theory.
}

\begin{document}

\maketitle

\newcommand{\ds}{\displaystyle}

\renewcommand{\H}{\mathcal{H}}
\newcommand{\br}{{\mbox{\boldmath$r$}}}
\newcommand{\bR}{{\mbox{\boldmath$R$}}}
\newcommand{\bS}{{\mbox{\boldmath$S$}}}
\newcommand{\bk}{{\mbox{\boldmath$k$}}}
\newcommand{\bH}{{\mbox{\boldmath$H$}}}
\newcommand{\bh}{{\mbox{\boldmath$h$}}}
\newcommand{\bJ}{{\mbox{\boldmath$J$}}}
\newcommand{\bI}{{\mbox{\boldmath$I$}}}
\newcommand{\bPsi}{{\mbox{\boldmath$\Psi$}}}
\newcommand{\bpsi}{{\mbox{\boldmath$\psi$}}}
\newcommand{\bPhi}{{\mbox{\boldmath$\Phi$}}}
\newcommand{\bd}{{\mbox{\boldmath$d$}}}
\newcommand{\bG}{{\mbox{\boldmath$G$}}}
\newcommand{\bu}{{\mbox{\boldmath$u$}}}
\newcommand{\be}{{\mbox{\boldmath$e$}}}
\newcommand{\bE}{{\mbox{\boldmath$E$}}}
\newcommand{\bp}{{\mbox{\boldmath$p$}}}
\newcommand{\bB}{{\mbox{\boldmath$B$}}}
\newcommand{\om}{{\omega_n}}
\newcommand{\omm}{{\omega_{n'}}}
\newcommand{\omd}{{\omega^2_n}}
\newcommand{\omt}{{\tilde{\omega}_{n}}}
\newcommand{\ommt}{{\tilde{\omega}_{n'}}}
\newcommand{\brho}{{\mbox{\boldmath$\rho$}}}
\newcommand{\bsigma}{{\mbox{\boldmath$\sigma$}}}
\newcommand{\bSigma}{{\mbox{\boldmath$\Sigma$}}}
\newcommand{\btau}{{\mbox{\boldmath$\tau$}}}
\newcommand{\bfeta}{{\mbox{\boldmath$\eta$}}}
\newcommand{\bskp}{{\mbox{\scriptsize\boldmath $k$}}}
\newcommand{\skp}{{\mbox{\scriptsize $k$}}}
\newcommand{\bsrp}{{\mbox{\scriptsize\boldmath $r$}}}
\newcommand{\bsRp}{{\mbox{\scriptsize\boldmath $R$}}}
\newcommand{\bsk}{\bskp}
\newcommand{\sk}{\skp}
\newcommand{\bsr}{\bsrp}
\newcommand{\bsR}{\bsRp}
\newcommand{\ri}{{\rm i}}
\newcommand{\re}{{\rm e}}
\newcommand{\rd}{{\rm d}}
\newcommand{\rM}{{\rm M}}
\newcommand{\rs}{{\rm s}}
\newcommand{\rt}{{\rm t}}
\newcommand{\Tc}{{$T_{\rm c}$}}
\renewcommand{\Pr}{{PrOs$_4$Sb$_{12}$}}
\newcommand{\La}{{LaOs$_4$Sb$_{12}$}}
\newcommand{\LaPr}{{(La$_{1-x}$Pr${_x}$)Os$_4$Sb$_{12}$}}
\newcommand{\PrLa}{{(Pr$_{1-x}$La${_x}$)Os$_4$Sb$_{12}$}}
\newcommand{\OsRu}{{Pr(Os$_{1-x}$Ru$_x$)$_4$Sb$_{12}$}}
\newcommand{\PrRu}{{PrRu$_4$Sb$_{12}$}}

\section{Introduction}
A crystal field quartet comprises fourfold degenerate electronic states represented by a
pseudospin-3/2 operator.
\cite{Santini09,Kuramoto09}
Unlike a pure spin, it shows anisotropic Zeeman splitting in the presence of a magnetic field.
\cite{Shiina97}
The origin of the anisotropy can be understood as an effect of octupole components produced
by the spin-3/2 operators, which are coupled to the magnetic field in the same way as dipole
components.
\par

For the spin-3/2 electronic states, the quadrupole degrees of freedom are coupled to strain
fields driven by external oscillating perturbations such as surface or bulk acoustic waves.
\cite{Hernandez-Minguez21,Vasselon23,Dietz23}
According to room temperature spin-acoustic resonance (SAR) measurements in
silicon carbide (4H-SiC),
\cite{Hernandez-Minguez20}
the electronic ground state of silicon vacancy (V$_{\rm Si}$) centers
can be described by a spin-3/2 quartet, and a characteristic of the quadrupole appears as an
anisotropic magnetoacoustic resonance (MAR) transition rate under the rotation of a static magnetic
field.
\cite{Koga24}
It is also natural that the quartet possesses octupole degrees of freedom.
\cite{Soltamov19}
So far, little attention has been paid to octupoles in the realization of the quantum control of
spin-3/2 quartets that will lead to an idea of operating such $V_{\rm Si}$ spins as qudits, namely,
four-level systems for quantum information and sensing applications.
\cite{Soltamov19,Kraus14,Widmann15,Simin17,Nagy18,Castelletto20,Son20}
Indeed, the Zeeman splitting of the $V_{\rm Si}$ quartet shows little magnetic anisotropy under a
strong field ($\sim$10~mT) and behaves like a pure spin-3/2.
\cite{Hernandez-Minguez20}
This indicates that the octupole effect is quite weak and hardly detected compared with
$f$-electron quartets with relatively strong spin-orbit coupling.
A typical example of the crystal field quartet is the $\Gamma_8$ representation of the cubic
point group $O_h$.
The evidence of octupole effects attributed to a $\Gamma_8$ quartet has been reported in
numerous studies of Ce-based $f$-electron systems such as CeB$_6$,
\cite{Shiina97,Sakai97,Shiina98,Shiba99,Mannix05,Kusunose05,Nagao06,Matsumura09,Nagao10,Matsumura12,Portnichenko20}
although it is still highly difficult to directly probe a characteristic of the octupole, which is usually
hidden by the $\Gamma_8$ magnetic properties, as well as that of the quadrupole.
\par

Motivated by the observation of the field-orientation-dependent SAR,
\cite{Hernandez-Minguez20}
we presented a more detailed theoretical description of the realistic spin--strain interaction
for spin-3/2 to quantitatively evaluate the spin--strain coupling parameters from the resonance
transition rate.
\cite{Koga24}
In this paper, we apply our previous theory to the crystal field quartet in a different way.
Unlike the isotropic spin-3/2, the energy differences between the quartet sublevels change with
the rotation of a magnetic field, so that the readjustment of a  resonance condition is inevitably
required.
To avoid this, we change the propagation direction of a standing surface acoustic wave (SAW) that
drives strain fields being coupled to the quartet, where the direction of the magnetic field is fixed at
a typical crystallographic axis.
For an $O_h$ $\Gamma_8$ quartet, we choose the field parallel to [110] or [111].
In actual MAR measurements, it is sufficient to adopt a few propagation directions of SAWs to 
obtain the information about (1) coupling parameters in the spin--strain (quadrupole--strain)
interaction and (2) an octupole effect associated with anisotropic Zeeman splitting of the quartet.
Indeed, the quantitative evaluation of these quantities is required for the quantum control of
solid-state spin multiplets considering the fine structure.
\cite{Soltamov19}
Thus, the method of MAR proposed here has the great advantage as a microscopic probe of both
quadrupole and octupole features possessed by the spin multiplets.
\par

In general, the multipole order phase depends on the direction of a magnetic field.
This is attributable to the anisotropic Zeeman splitting, as mentioned above, which can be
represented by the octupole components classified into the point group representations.
The octupole effect is significant for the $\Gamma_8$ multipole--multipole interaction systems
having a nearly SU(4) symmetry, such as CeB$_6$.
\cite{Shiina97}
On the basis of a spherical interaction model, we demonstrate how to identify a quadrupole order
parameter depending on the field direction using MAR.
\par

Currently, the frequency of an SAW-induced strain field is limited up to gigahertz order.
Usually, a relatively strong magnetic field is required for precise measurements, and
a necessary resonance condition for MAR possibly exceeds ten gigahertz order in energy.
To realize the MAR in large excitation energy gaps, we propose a possible method combining a
microwave with strain fields.
\cite{Koga20a,Matsumoto20,Koga20b,Koga22}
The difficulty can be resolved by a high-frequency photon absorption transition using
a linearly polarized microwave, which is accompanied by a simultaneous low-frequency phonon
absorption.
The total transition process is analogous with bichromatic driving for electron spin resonance
(ESR) measurements.
\cite{Gromov00,Kalin04,Gyorgy22}
On the basis of the Floquet theory,
\cite{Shirley65,Son09}
we formulate the hybrid MAR as a useful method for the evaluation of the quadrupole-coupling
strengths associated with the phonon absorption.
\par

This paper is organized as follows.
In Sect.~2, we describe the multipole degrees of freedom for a $\Gamma_8$ quartet
and the anisotropic Zeeman splitting derived from the octupoles under the rotation of a magnetic field.
In Sect.~3, we formulate a theory of acoustically driven resonance transition depending on the
SAW propagation directions considering the quadrupole--strain interactions under the typical field
directions.
How to evaluate an octupole effect is explained, and lastly, our proposal for the hybrid MAR
measurements is presented.
The conclusion of this paper is given in Sect.~4.
In Appendix~A, we list the multipole operators with the $4 \times 4$ matrix form.
The equations listed in Appendices~B and C are used for the anisotropic Zeeman and
quadrupole--strain coupling terms, respectively, under the field rotation.
In Appendix~D, the multipole operators are represented by pseudospins for spin and orbital
degrees of freedom in the quartet.
To discuss anisotropic magnetic field dependences arising from the Zeeman splitting in a
quadrupole order phase, we introduce a multipole--multipole interaction model in Appendix~E,
and demonstrate how the octupole effects are derived from the Ginzburg--Landau (GL) expansion
of the free energy for the typical field directions in Appendix~F.

\section{Magnetic Field Effect on a Crystal Field Quartet}
\subsection{Multipole degrees of freedom}
To investigate multipole degrees of freedom in a crystal field quartet, we consider the following
$J = 5/2$ $\Gamma_8$ states for the cubic point group $O_h$: 
\begin{align}
& | \Gamma_{8,1} \rangle
= - \sqrt{ \frac{1}{6} } | \frac{3}{2} \rangle - \sqrt{ \frac{5}{6} } | - \frac{5}{2} \rangle,
\nonumber \\
& | \Gamma_{8,2} \rangle = | \frac{1}{2} \rangle,
\nonumber \\
& | \Gamma_{8,3} \rangle = - | - \frac{1}{2} \rangle,
\nonumber \\
& | \Gamma_{8,4} \rangle
= \sqrt{ \frac{5}{6} } | \frac{5}{2} \rangle + \sqrt{ \frac{1}{6} } | - \frac{3}{2} \rangle,
\end{align}
where $| M \rangle$ ($M = 5/2, 3/2, \cdots, - 5/2$) represents one of the $J_z$ states with total
angular momentum $J = 5/2$.
The multipoles of the $J = 5/2$ moment are described by tensor operators, which are expressed
by the products of the angular momentum operators $\{ J_x, J_y, J_z \}$.
Because we here restrict ourselves to multipoles in the $\Gamma_8$ quartet,
the irreducible tensor operators for the $\Gamma_8$ basis can be expressed by effective
$J = 3/2$ dipole (rank-1), quadrupole (rank-2), and octupole (rank-3) operators.
The irreducible tensor operators $J_q^{(p)}$ of rank $p$ ($q = p, p - 1, \cdots, -p$) are constructed
using the following formula:
\cite{Inui90}
\begin{align}
& J_p^{(p)} = (- 1)^p \sqrt{ \frac{ (2p - 1)(2p - 3) \cdots 3 \cdot 1}{2p(2p - 2) \cdots 2} } J_+^p, \\
& J_{q - 1}^{(p)} = \frac{1}{ \sqrt{ (p + q)(p - q + 1) } } [ J_-, J_q^{(p)} ],
\end{align}
and $J_\pm = J_x \pm i J_y$.
In Appendix~A, we list the three dipole, five quadrupole, and seven octupole operators for $O_h$,
and show the corresponding $4 \times 4$ matrices for $J = 3/2$.
\par

Let us represent the $\Gamma_8$ quartet by $J_{\Gamma_8} = 3/2$ pseudospin as
$\{ | \Gamma_{8,1} \rangle,  | \Gamma_{8,2} \rangle,  | \Gamma_{8,3} \rangle,
| \Gamma_{8,4} \rangle \}
\rightarrow \{ | 3/2 \rangle, | 1/2 \rangle, | - 1/2 \rangle, | - 3/2 \rangle \}$.
The matrix expressions of $J_{\Gamma_8} = 3/2$ dipole operators are obtained by calculating the matrix
elements of $J = 5/2$ dipole operators in the $\Gamma_8$ subspace as
\begin{align}
& J_{x,\Gamma_8} =
\left(
\begin{array}{cccc}
0 & \ds{ - \frac{1}{ \sqrt{3} } } & 0 & \ds{ - \frac{5}{6} } \\
\ds{ - \frac{1}{ \sqrt{3} } } & 0 & \ds{ - \frac{3}{2} } & 0 \\
0 & \ds{ - \frac{3}{2} } & 0 & \ds{ - \frac{1}{ \sqrt{3} } }  \\
\ds{ - \frac{5}{6} } & 0 & \ds{ - \frac{1}{ \sqrt{3} } } & 0
\end{array}
\right),
\nonumber \\
& J_{y,\Gamma_8} =
\left(
\begin{array}{cccc}
0 & \ds{ \frac{i}{ \sqrt{3} } } & 0 & \ds{ -i \frac{5}{6} } \\
\ds{ - \frac{i}{ \sqrt{3} } } & 0 & \ds{ i \frac{3}{2} } & 0 \\
0 & \ds{ -i \frac{3}{2} } & 0 & \ds{ \frac{i}{ \sqrt{3} } }  \\
\ds{ i \frac{5}{6} } & 0 & \ds{ - \frac{i}{ \sqrt{3} } } & 0
\end{array}
\right),
\nonumber \\
& J_{z,\Gamma_8} =
\left(
\begin{array}{cccc}
\ds{ - \frac{11}{6} } & 0 & 0 & 0 \\
0 & \ds{ \frac{1}{2} } & 0 & 0 \\
0 & 0 & \ds{ - \frac{1}{2} } & 0 \\
0 & 0 & 0 & \ds{ \frac{11}{6} }
\end{array}
\right).
\label{eqn:G8dipole}
\end{align}
With Eqs.~(\ref{eqn:Jmu}) and (\ref{eqn:Talpha}), the dipole components in
Eq.~(\ref{eqn:G8dipole}) are represented by the linear combination of $J = 3/2$ dipole
$J_\mu$ and octupole $T_\mu^\alpha$ operators:
$J_{\mu,\Gamma_8} = - J_\mu - (4/9) T_\mu^\alpha$ ($\mu = x, y, z$).
The spin-orbit coupling brings about anisotropic Zeeman splitting described by
$T_\mu^\alpha$ in $J_{\mu,\Gamma_8}$.
To investigate the octupole effect, we generalize $J_{\mu, \Gamma_8}$
by introducing an anisotropy parameter $\lambda$ as the coefficient of $T_\mu^\alpha$
in the following discussion.
For a large $\lambda$, the Zeeman splitting strongly depends on the field direction.
The Zeeman term is then given by the following Hamiltonian,
\begin{align}
H_Z =  g_J \mu_B \sum_\mu J_{\mu,\Gamma_8} H_\mu
= g_J \mu_B \sum_\mu ( - J_\mu - \lambda T_\mu^\alpha ) H_\mu~~(\mu = x, y, z),
\label{eqn:HZ}
\end{align}
with the external magnetic field $\bH = (H_x, H_y, H_z)$.
Here, $\mu_B$ is the Bohr magneton and $g_J$ is Land{\' e}'s g factor ($g_J = 6/7$ for localized
$f$-electrons with $J = 5/2$).

\subsection{Zeeman splitting under field rotation}
The cubic reference frame is defined by the coordinates $\{ x, y, z \}$ with the three orthogonal
vectors $\be_x = (1, 0, 0)$, $\be_y = (0, 1, 0)$, and $\be_z = (0, 0, 1)$.
To describe the Zeeman splitting under the magnetic field $\bH$, it is convenient to use a different
reference frame where the field direction is chosen as the quantization axis $Z$.
The polar representation of the field is written as $\bH = H \be_Z$ with
$\be_Z = ( \sin \theta \cos \phi, \sin \theta \sin \phi, \cos \theta )$;
the other orthogonal unit vectors are defined as
$\be_X = ( - \cos \theta \cos \phi, - \cos \theta \sin \phi, \sin \theta )$ and
$\be_Y = ( \sin \phi, - \cos \phi, 0)$.
In the new ($XYZ$) reference frame,  we define the following unitary transformation:
$J_\nu = U^\dagger ( \be_\nu \cdot \bJ ) U$ ($\nu = X, Y, Z$).
Here, the operator $U$ is given by the $4 \times 4$ matrix form in Appendix~B.
The dipole term in Eq.~(\ref{eqn:HZ}) is then transformed as
\begin{align}
U^\dagger \left( \sum_\mu J_\mu  B_\mu \right) U = J_Z B~~
( B_\mu \equiv g_J \mu_B H_\mu, B \equiv g_J \mu_B H),
\label{eqn:UJB}
\end{align}
while the transformed octupole term becomes very complicated.
After a lengthy calculation, we obtain
\begin{align}
& U^\dagger \left( \sum_\mu T_\mu^\alpha B_\mu \right) U = f_1 T_Z^\alpha + f_2 T_Z^\beta + f_3 T_{XYZ}
\nonumber \\
&~~~~~~~~~~~~~~
+ f_4 \left( \sqrt{ \frac{3}{8} } T_X^\alpha + \sqrt{ \frac{5}{8} } T_X^\beta \right)
+ f_5 \left( \sqrt{ \frac{5}{8} } T_X^\alpha - \sqrt{ \frac{3}{8} } T_X^\beta \right)
\nonumber \\
&~~~~~~~~~~~~~~
+f_6 \left( \sqrt{ \frac{3}{8} } T_Y^\alpha - \sqrt{ \frac{5}{8} } T_Y^\beta \right)
+f_7 \left( - \sqrt{ \frac{5}{8} } T_Y^\alpha - \sqrt{ \frac{3}{8} } T_Y^\beta \right),
\label{eqn:UTB}
\end{align}
where the coefficients $f_i$ ($i = 1, 2, \cdots, 7$) are given in Eq.~(\ref{eqn:TBf}).
The Zeeman splitting is described by the transformed Hamiltonian
\begin{align}
\tilde{H}_Z = U^\dagger \sum_\mu ( - J_\mu  - \lambda T_\mu^\alpha ) B_\mu U
\end{align}
in the ($XYZ$) reference frame.
Owing to the cubic symmetry, the octupole part is reduced to simpler forms in the specific field
directions $\bH \parallel [110]$ ($\cos \theta = 0$, $\sin \theta = 1$, $\phi = \pi / 4$) and
$\bH \parallel [111]$ ($\cos \theta = \sqrt{ 1/3 }$, $\sin \theta = \sqrt{ 2/3 }$, $\phi = \pi / 4$).
For $\bH \parallel [110]$, the explicit matrix form is given by
\begin{align}
& \tilde{H}_{Z, [110]} = B \left[ - J_Z
- \lambda \left( - \frac{1}{4} T_Z^\alpha - \frac{ \sqrt{15} }{4} T_Z^\beta \right) \right]
\nonumber \\
&~~~~~~
= - \frac{B}{16} \left(
\begin{array}{cccc}
24 - 3 \lambda & 0 & - 15 \sqrt{3} \lambda & 0 \\
0 & 8 + 9 \lambda & 0 & 15 \sqrt{3} \lambda \\
- 15 \sqrt{3} \lambda & 0 & - 8 - 9 \lambda & 0 \\
0 & 15 \sqrt{3} \lambda & 0 & - 24 + 3 \lambda
\end{array}
\right).
\label{eqn:HZ110}
\end{align}
For $\bH \parallel [111]$,
\begin{align}
& \tilde{H}_{Z, [111]} = B \left\{ - J_Z
- \lambda \left[ - \frac{2}{3} T_Z^\alpha
+ \frac{ \sqrt{5} }{3} \left( \sqrt{ \frac{5}{8} } T_X^\alpha - \sqrt{ \frac{3}{8} } T_X^\beta \right) \right]
\right\}
\nonumber \\
&~~~~~~
= - \frac{B}{4} \left(
\begin{array}{cccc}
6 - 2 \lambda & 0 & 0 & 5 \sqrt{2} \lambda \\
0 & 2 + 6 \lambda & 0 & 0 \\
0 & 0 & - 2 - 6 \lambda & 0 \\
5 \sqrt{2} \lambda & 0 & 0 & - 6 + 2 \lambda
\end{array}
\right).
\label{eqn:HZ111}
\end{align}
Diagonalizing these matrices, we obtain the eigenvalues $E_{Z,i}$ ($i = 1, 2, 3, 4$) and the
corresponding eigenstates $| \psi_i \rangle$ as listed in Table~I.
\begin{table}
\caption{
Eigenenergies normalized by $B = g_J \mu_B H$ and the corresponding eigenstates of a crystal
field quartet split by the magnetic fields $\bH \parallel [110]$ and $\bH \parallel [111]$.
Each eigenstate is described by the $J = 3/2$ states $| M \rangle$ ($M = 3/2, 1/2, -1/2, -3/2$),
which are quantized with respect to the field direction.
The anisotropic parameter $\lambda$ represents an octupole effect on the Zeeman splitting.
}
\begin{center}
\begin{tabular}{lll} \hline
$E_{Z, i}$ for $\bH \parallel [110]$ & $| \psi_i \rangle$ \\ \hline
~\vspace{-0.3cm} \\
$\ds{ E_{Z,1} =  - \frac{1}{8} \left( 4 - 3 \lambda + \frac{F_\lambda}{2} \right) }$
& $\ds{ | \psi_1 \rangle = c_+ | \frac{3}{2} \rangle + c_- | - \frac{1}{2} \rangle }$ \\
~\vspace{-0.3cm} \\
$\ds{ E_{Z, 2} =  - \frac{1}{8} \left( - 4 + 3 \lambda + \frac{F_\lambda}{2} \right) }$
& $\ds{ | \psi_2 \rangle = - c_- | - \frac{3}{2} \rangle + c_+ | \frac{1}{2} \rangle }$ \\
~\vspace{-0.3cm} \\
$\ds{ E_{Z, 3} =  \frac{1}{8} \left( - 4 + 3 \lambda + \frac{F_\lambda}{2} \right) }$
& $\ds{ | \psi_3 \rangle = - c_- | \frac{3}{2} \rangle + c_+ | - \frac{1}{2} \rangle }$ \\
~\vspace{-0.3cm} \\
$\ds{ E_{Z, 4} =  \frac{1}{8} \left( 4 - 3 \lambda + \frac{F_\lambda}{2} \right) }$
& $\ds{ | \psi_4 \rangle = c_+ | - \frac{3}{2} \rangle + c_- | \frac{1}{2} \rangle }$ \\
~\vspace{-0.3cm} \\ \hline \hline
~\vspace{-0.3cm} \\
$F_\lambda = \sqrt{ ( 16 + 3 \lambda )^2 + ( 15 \sqrt{3} \lambda )^2 }$,
& $\ds{ c_\pm = \pm \sqrt{ \frac{1}{2} \left( 1 \pm \frac{ 16 + 3 \lambda }{F_\lambda} \right) } }$
\\ \hline 
\end{tabular}
\end{center}
~ \\
\begin{center}
\begin{tabular}{lll} \hline
$E_{Z, i}$ for $\bH \parallel [111]$ & $| \psi_i \rangle$ \\ \hline
~\vspace{-0.3cm} \\
$\ds{ E_{Z, 1} =  - \frac{1}{4} G_\lambda }$
& $\ds{ | \psi_1 \rangle = d_+ | \frac{3}{2} \rangle + d_- | - \frac{3}{2} \rangle }$ \\
~\vspace{-0.3cm} \\
$\ds{ E_{Z, 2} =  - \frac{1}{2} ( 1 + 3 \lambda ) }$
& $\ds{ | \psi_2 \rangle = | \frac{1}{2} \rangle }$ \\
~\vspace{-0.3cm} \\
$\ds{ E_{Z, 3} =  \frac{1}{2} ( 1 + 3 \lambda ) }$
& $\ds{ | \psi_3 \rangle = | - \frac{1}{2} \rangle }$ \\
~\vspace{-0.3cm} \\
$\ds{ E_{Z, 4} =  \frac{1}{4} G_\lambda }$
& $\ds{ | \psi_4 \rangle = - d_- | \frac{3}{2} \rangle + d_+ | - \frac{3}{2} \rangle }$ \\
~\vspace{-0.3cm} \\ \hline \hline
~\vspace{-0.3cm} \\
$G_\lambda = \sqrt{ ( 6 - 2 \lambda )^2 + 50 \lambda^2 }$,
& $\ds{ d_\pm = \sqrt{ \frac{1}{2} \left( 1 \pm \frac{ 6 - 2 \lambda }{G_\lambda} \right) } }$ \\ \hline 
\end{tabular}
\end{center}
\end{table}

\section{Acoustically Driven Resonance Transition}
\subsection{Quadrupole--strain interaction under the field rotation}
In the local $O_h$ systems, the quadrupole--strain interaction is described by the following
Hamiltonian:
\cite{Nakamura94}
\begin{align}
H_\varepsilon = g_3 ( O_u \varepsilon_u + O_v \varepsilon_v )
+ g_5 ( O_{zx} \varepsilon_{zx} + O_{xy} \varepsilon_{xy} + O_{yz} \varepsilon_{yz} ),
\label{eqn:Hep0}
\end{align}
with the two independent coupling constants $g_3$ and $g_5$.
Each quadrupole is coupled to the corresponding strain tensor $\varepsilon_{\mu \nu}$
($\mu, \nu = x, y, z$) and
\begin{align}
\varepsilon_u \equiv \frac{1}{ \sqrt{3} } ( 2 \varepsilon_{zz} - \varepsilon_{xx} - \varepsilon_{yy} ),~~
\varepsilon_v \equiv \varepsilon_{xx} - \varepsilon_{yy}.
\end{align}
For the magnetic field $\bH \parallel Z$, we apply the unitary transformation to
$H_\varepsilon$ using Eq.~(\ref{eqn:US}) and obtain its form in the $(XYZ)$ reference frame as
\cite{Koga24}
\begin{align}
\tilde{H}_\varepsilon = U^\dagger H_\varepsilon U = \sum_K A_K O_K,
\label{eqn:Hep}
\end{align}
where $O_K$ ($K = U, V, ZX, XY, YZ$) are the quadrupole operators in the ($XYZ$) reference
frame.
As shown in Appendix~C, each unitary transformed $O_k$ is given by the linear combination of
the five components $O_K$.
Accordingly, the strain-dependent coupling coefficients $A_K$ depend on the field direction
represented by $\theta$ and $\phi$.

\subsection{Single-phonon transition rate depending on propagation directions of SAW}
\begin{figure}
\begin{center}
\includegraphics[width=7cm,clip]{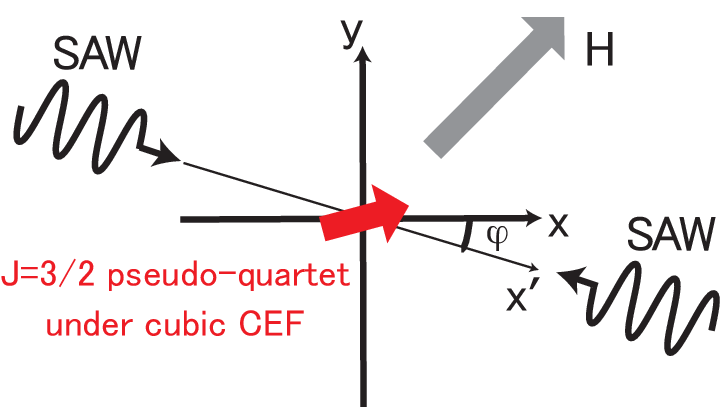}
\end{center}
\caption{
(Color online)
Illustration of pseudo-quartet represented by $J = 3/2$ operator in a cubic crystalline electric field
(CEF) environment, which is coupled to SAWs propagating in the $\pm x'$ directions under the
static magnetic field $\bH$.
The angle $\varphi$ of the $x'$-axis is measured from the crystallographic $x$-axis.
}
\label{fig:1}
\end{figure}
We consider that the local quartet is coupled to the strain fields driven by a plane Rayleigh SAW
propagating on the $xy$ plane, as shown in Fig.~\ref{fig:1}.
The strain tensors are described by
\begin{align}
\varepsilon_{\alpha \beta} (t, x', z) = \varepsilon_{\alpha \beta} (z) e^{ i k x' - i \omega t }
+ \varepsilon_{\alpha \beta}^* (z) e^{ -i k x' + i \omega t },
\label{eqn:SAW}
\end{align}
where the $x'$-axis denotes the direction of the propagating SAW, nonzero components
$\alpha \beta$ are restricted to $\{ x'x', zz, x'z \}$ (no displacement along $y'$), and the amplitudes
of strain tensors only depend on the depth $z$ from the surface layer ($xy$ plane).
We assume that $\varepsilon_{x'x'}$ and $\varepsilon_{zz}$ are real, whereas
$\varepsilon_{x'z} \equiv - i \varepsilon''_{x'z}$ is purely imaginary ($\varepsilon''_{x'z}$ is real).
\cite{Hernandez-Minguez20}
This indicates an elliptical polarization of the SAW medium.
When the $x'$-axis is rotated from $x$ by angle $\varphi$ in the $xy$ plane, the coordinates are
transformed as
\cite{Koga24}
\begin{align}
\left(
\begin{array}{c}
x' \\
y'
\end{array}
\right)
= \left(
\begin{array}{cc}
\cos \varphi & \sin \varphi \\
- \sin \varphi & \cos \varphi
\end{array}
\right)
\left(
\begin{array}{c}
x \\
y
\end{array}
\right),
\end{align}
and the strain tensor components in Eq.~(\ref{eqn:AK}) are replaced as
\begin{align}
& \varepsilon_{xx} \rightarrow \varepsilon_{x' x'} \cos^2 \varphi,~~
\varepsilon_{yy} \rightarrow \varepsilon_{x' x'} \sin^2 \varphi,~~
\varepsilon_{xy} \rightarrow \varepsilon_{x'x'} \sin \varphi \cos \varphi,
\nonumber \\
& \varepsilon_{zx} \rightarrow \varepsilon_{x' z} \cos \varphi,~~
\varepsilon_{yz} \rightarrow \varepsilon_{x' z} \sin \varphi.
\label{eqn:newstrain}
\end{align}
Here, $\varepsilon_{\alpha \beta} = 0$ when either $\alpha$ or $\beta$ is $y'$.
\par

From now on, we restrict ourselves to the two cases of field direction in Table~I to
calculate the transition matrix element between two quartet states
$M_{mn, \pm} = \langle \psi_m | \tilde{H}_\varepsilon | \psi_n \rangle$
using Eq.~(\ref{eqn:Hep}), where the transition is caused by the absorption of a single phonon.
The subscripts $+$ and $-$ denote $+x'$ and $-x'$ propagation directions of the SAW,
respectively.
For the latter, $k$ is replaced by $-k$ in Eq.~(\ref{eqn:SAW}), and
$\varepsilon_{x'z}'' \rightarrow - \varepsilon_{x'z}''$ is required.
For $\bH \parallel [110]$, we obtain the transition matrix elements associated with the ground state
$| \psi_1 \rangle$:
\begin{align}
& M_{21,+} = i \sqrt{3} \left[ g_3 \varepsilon_{x'x'} \cos 2 \varphi
- g_5 \varepsilon_{x'z}'' \cos \left( \varphi - \frac{ \pi }{4} \right) \right],
\nonumber \\
& M_{31,+} = \left[ c_+ c_- + \frac{ \sqrt{3} }{2} ( c_+^2 - c_-^2 ) \right]
g_3 ( 2 \varepsilon_{zz} - \varepsilon_{x'x'} )
\nonumber \\
&~~~~~~~~
+ \left[ -\frac{3}{2} c_+ c_- + \frac{ \sqrt{3} }{4} ( c_+^2 - c_-^2 ) \right] g_5 \varepsilon_{x'x'} \sin 2 \varphi
\nonumber \\
&~~~~~~~~
+ \sqrt{3} g_5 \varepsilon_{x'z}'' \cos \left( \varphi + \frac{ \pi }{4} \right),
\nonumber \\
& M_{41,+} = 0,
\end{align}
and $M_{mn,+} = M_{nm,+}^*$.
In the same manner, for $\bH \parallel [111]$,
\begin{align}
& M_{21,+} = \left( \sqrt{ \frac{2}{3} } d_+ + \frac{ d_- }{ \sqrt{3} } \right)
g_3 ( 2 \varepsilon_{zz} - \varepsilon_{x'x'} )
\nonumber \\
&~~~~~~~~
+ i \left( \sqrt{2} d_+ + d_- \right) g_3 \varepsilon_{x'x'} \cos 2 \varphi
\nonumber \\
&~~~~~~~~
+ \left( - \frac{ d_+ }{ \sqrt{6} } + \frac{ d_- }{ \sqrt{3} } \right) g_5 \varepsilon_{x'x'} \sin 2 \varphi
\nonumber \\ 
&~~~~~~~~
+ \left( d_+ - \sqrt{2} d_- \right)
g_5 \varepsilon_{x'z}'' \cos \left( \varphi + \frac{ \pi }{4} \right)
\nonumber \\
&~~~~~~~~
+ i \left( - \frac{ d_+ }{ \sqrt{3} } + \sqrt{ \frac{2}{3} } d_- \right)
g_5 \varepsilon_{x'z}'' \cos \left( \varphi - \frac{ \pi }{4} \right),
\nonumber \\
& M_{31,+} = \left( - \sqrt{ \frac{2}{3} } d_- + \frac{ d_+ }{ \sqrt{3} } \right)
g_3 ( 2 \varepsilon_{zz} - \varepsilon_{x'x'} )
\nonumber \\
&~~~~~~~~
+ i \left( \sqrt{2} d_- - d_+ \right) g_3 \varepsilon_{x'x'} \cos 2 \varphi
\nonumber \\
&~~~~~~~~
+ \left( \frac{ d_- }{ \sqrt{6} } + \frac{ d_+ }{ \sqrt{3} } \right) g_5 \varepsilon_{x'x'} \sin 2 \varphi
\nonumber \\ 
&~~~~~~~~
+ \left( d_- + \sqrt{2} d_+ \right)
g_5 \varepsilon_{x'z}'' \cos \left( \varphi + \frac{ \pi }{4} \right)
\nonumber \\
&~~~~~~~~
+ i \left( \frac{ d_- }{ \sqrt{3} } + \sqrt{ \frac{2}{3} } d_+ \right)
g_5 \varepsilon_{x'z}'' \cos \left( \varphi - \frac{ \pi }{4} \right),
\nonumber \\
& M_{41,+} = 0.
\end{align}
\par

Now, we are prepared to calculate the single-phonon transition rate $W_{n1}$ between the
ground and $n$th excited states of the quartet.
We here consider the case of a standing SAW where the quartet is coupled simultaneously to
two SAWs propagating along the $+x'$ and $-x'$ directions with equal intensities.
Indeed, a standing wave was realized in a recent spin acoustic resonance experiment.
\cite{Hernandez-Minguez20}
Thus, the rate $W_{n1}$ is proportional to the sum of the $\pm x'$ propagating SAW contributions,
\begin{align}
W_{n1} \propto \frac{1}{2} \langle | M_{n1,+} |^2 + | M_{n1,-} |^2 \rangle,
\end{align}
where the bracket $\langle \cdots \rangle$ indicates averaging the strain amplitudes with
respect to the spatial distributions of magnetic ions with the quartets.
We have also assumed that the coherence terms such as
$\langle {\rm Re} [ M_{n1,+}^* M_{n1,-} ] \cdot \cos 2 k x' \rangle$ are canceled out by the average
over the ion positions.
\par

\begin{figure}
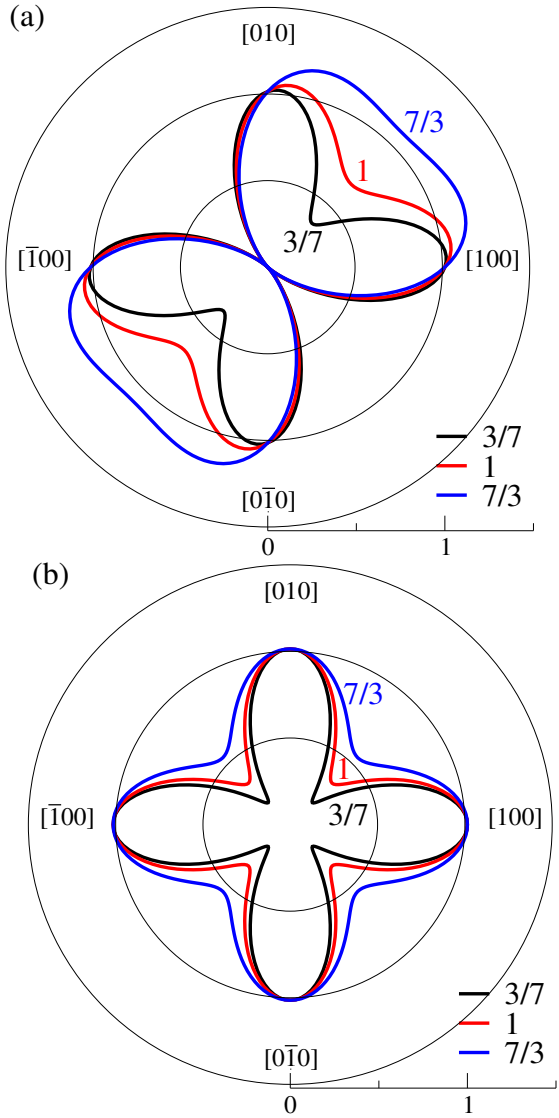

\begin{center}
\includegraphics[width=7cm,clip]{fig2a.eps}
\hspace*{0.5cm}
\includegraphics[width=7cm,clip]{fig2b.eps}
\end{center}
\caption{
(Color online)
Polar representation of transition rate for $\bH \parallel [110]$ with respect to the angle
$\varphi$ of the SAW propagation direction in Eq.~(\ref{eqn:Wn1}).
The data plotted here are normalized by the values at $\varphi = 0$.
(a)  Plotting of $W_{21}^{[110]} (\varphi)$ for
$g_5^2 \langle \varepsilon_{x'z}''^2 \rangle /  g_3^2 \langle \varepsilon_{x'x'}^2 \rangle = 3/7$
(black), $1$ (red), and $7/3$ (blue).
(b) $W_{21}^{[110]} (\varphi) + W_{21}^{[110]} (- \varphi)$ for evaluating the amplitude
$C_{21}^{[110]}$ of the $\cos 4 \varphi$ term.
}
\label{fig:2}
\end{figure}
Under the magnetic field $\bH$, the rate $W_{n1}^H$ is calculated as a function of
$\varphi$:
\begin{align}
W_{n1}^H ( \varphi ) \propto A_{n1}^H + B_{n1}^H \sin 2 \varphi + C_{n1}^H \cos 4 \varphi.
\label{eqn:Wn1}
\end{align}
For $\bH \parallel [110]$, the coefficients in $W_{21}^H$ are given by
\begin{align}
& A_{21}^{[110]} = \frac{1}{2} \left( g_3^2 \langle \varepsilon_{x'x'}^2 \rangle
+ g_5^2 \langle \varepsilon_{x'z}''^2 \rangle \right),
\nonumber \\
& B_{21}^{[110]} = \frac{1}{2} g_5^2 \langle \varepsilon_{x'z}''^2 \rangle,
\nonumber \\
& C_{21}^{[110]} = \frac{1}{2} g_3^2 \langle \varepsilon_{x'x'}^2 \rangle.
\label{eqn:ABC110}
\end{align}
Here, $\langle \cdots \rangle$ represents an average over the $z$ direction.
In Fig.~\ref{fig:2}(a), the polar representation of $W_{21} ( \varphi )$ is plotted for various values
of the ratio
$g_5^2 \langle \varepsilon_{x'z}''^2 \rangle / g_3^2 \langle \varepsilon_{x'x'}^2 \rangle$.
As the ratio increases, the term $\sin 2 \varphi$ becomes more dominant than $\cos 4 \varphi$,
and the latter is characterized by a V-shaped feature at $\varphi = \pi / 4$ that matches the field
direction.
Similarly, for the rate $W_{31}^{[110]}$,
\begin{align}
& A_{31}^{[110]} = ( \alpha_3 g_3 )^2 \langle ( 2 \varepsilon_{zz} - \varepsilon_{x'x'} )^2 \rangle
+ \frac{1}{2} g_5^2 ( \alpha_5^2 \langle \varepsilon_{x'x'}^2 \rangle
+ \langle \varepsilon_{x'z}''^2 \rangle ),
\nonumber \\
& B_{31}^{[110]} = 2 \alpha_3 \alpha_5 g_3 g_5
\langle ( 2 \varepsilon_{zz} - \varepsilon_{x'x'} ) \varepsilon_{x'x'} \rangle
- \frac{1}{2} g_5^2 \langle \varepsilon_{x'z}''^2 \rangle,
\nonumber \\
& C_{31}^{[110]} = - \frac{1}{2} ( \alpha_5 g_5 )^2 \langle \varepsilon_{x'x'}^2 \rangle,
\end{align}
where
\begin{align}
\alpha_3 = \frac{ c_+ c_- }{ \sqrt{3} } + \frac{ c_+^2 - c_-^2 }{2},~~
\alpha_5 = -\frac{ \sqrt{3} }{2} c_+ c_- + \frac{ c_+^2 - c_-^2 }{4}.
\end{align}
In particular, we focus on $C_{n1}^H$ in the term of $\cos 4 \varphi$ to evaluate the
ratio of coupling constants $r_g \equiv g_5 / g_3$ in Eq.~(\ref{eqn:Hep0}) because it contains only
$\langle \varepsilon_{x'x'}^2 \rangle$ among various strain components.
For this purpose, the contribution from the term of $\cos 4 \varphi$ is extracted by plotting
$W_{n1}^H ( \varphi ) + W_{n1}^H ( - \varphi )$, as shown in Fig.~\ref{fig:2}(b).
Indeed, the value of $C_{n1}^H$ can be obtained from measurable values of $W_{n1}^H$ as
\begin{align}
C_{n1}^H \propto 2 W_{n1}^H (0) - W_{n1}^H ( \pi / 4 ) - W_{n1}^H ( - \pi /4 ).
\end{align}
From Table~I, $\alpha_5$ is given as a function of $\lambda$ representing the octupole
contribution, which leads to
\begin{align}
- \frac{ C_{31}^{[110]} }{ C_{21}^{[110]} } = \alpha_5^2 r_g^2
= \left( \frac{ g_5 }{ g_3} \right)^2
\frac{ 16 ( 1 + 3 \lambda )^2 }{ ( 16 + 3 \lambda )^2 + ( 15 \sqrt{3} \lambda )^2 }.
\end{align}
A similar argument can be applied to $W_{n1}^{[111]}$ for $\bH \parallel [111]$.
Because $C_{n1}^{[111]}$ ($n = 2, 3$) contains the only parameter
$\langle \varepsilon_{x'x'}^2 \rangle$ for strain as  $C_{21}^{[110]}$ in Eq.~(\ref{eqn:ABC110}),
we obtain
\begin{align}
& \frac{ C_{21}^{[111]} }{ C_{21}^{[110]} } =
\frac{1}{3} \left[ ( \sqrt{2} d_+ + d_- )^2 - \frac{ r_g^2 }{6} ( d_+ - \sqrt{2} d_- )^2 \right]
\nonumber \\
&~~~~~~~~~~
= \left( \frac{1}{2} + \frac{ 1 + 3 \lambda }{G_\lambda} \right)
- \frac{ r_g^2 }{6} \left( \frac{1}{2} - \frac{ 1 + 3 \lambda }{G_\lambda} \right),
\nonumber \\
& \frac{ C_{31}^{[111]} }{ C_{21}^{[110]} } =
\frac{1}{3} \left[ ( d_+ - \sqrt{2} d_-  )^2 - \frac{ r_g^2 }{6} ( \sqrt{2} d_+ + d_-  )^2 \right]
\nonumber \\
&~~~~~~~~~~
= \left( \frac{1}{2} - \frac{ 1 + 3 \lambda }{G_\lambda} \right)
- \frac{ r_g^2 }{6} \left( \frac{1}{2} + \frac{ 1 + 3 \lambda }{G_\lambda} \right),
\end{align}
where $G_\lambda$ is given as a function of $\lambda$ in Table~I.
Combining these equations, we can evaluate the coupling-constant ratio as
\begin{align}
\left( \frac{ g_5 }{ g_3 } \right)^2
= 6 \left( 1 - \frac{ C_{21}^{[111]} + C_{31}^{[111]} }{ C_{21}^{[110]} } \right).
\end{align}
Suppose the parameter $\lambda$ is unknown, it can also be evaluated using $C_{n1}^H$ as
\begin{align}
\frac{ G_\lambda + 2 ( 1 + 3 \lambda ) }{ G_\lambda - 2 ( 1 + 3 \lambda ) } =
\frac{ C_{31}^{[111]} - C_{21}^{[110]} }{ C_{21}^{[111]} - C_{21}^{[110]} }.
\end{align}
For the $J = 5/2$ $\Gamma_8$ quartet ($\lambda = 4/9$), the calculated value of this ratio equals
eight, which is four times larger than that for the pure $J = 3/2$ quartet ($\lambda = 0$).

\subsection{Anisotropy of Zeeman effect on quadrupole order phase}
It is highly useful to study the octupole effect in the antiferroquadrupole order phase on the basis
of the multipole--multipole interaction model given in Appendix~E, where we use pseudospin
representations for the multipole operators, as introduced in Appendix~D.
The detailed group theoretical study was first presented by Shiina {\it et al.},
\cite{Shiina97}
and an important role of octupole degrees of freedom in CeB$_6$ was elucidated.
The spherical interaction Hamiltonian comprises the nonmagnetic part $H_{\rm Q}$ with a
coupling constant $D_{\rm Q}$ in Eq.~(\ref{eqn:HQ}) and the magnetic part $H_{\rm M}$ with
$D_{\rm M}$ in Eq.~(\ref{eqn:HM}).
Note that the forms of both $H_{\rm Q}$ and $H_{\rm M}$ are invariant in arbitrary
directions of the magnetic field $\bH$.
This also holds for $D_{\rm Q} \ne D_{\rm M}$ in the combined Hamiltonian of $H_{\rm Q}$ and
$H_{\rm M}$, where the symmetry of the interaction is lowered from the SU(4) symmetry
satisfying $D_{\rm Q} = D_{\rm M}$ in Eq.~(\ref{eqn:Hs}).
Thus, under the rotation of $\bH$, the model Hamiltonian
$\tilde{H}_s = \tilde{H}_{\rm Q} + \tilde{H}_{\rm M} + \tilde{H}_{\rm Z}$  in the $(XYZ)$ reference
frame can be written as
\begin{align}
& \tilde{H}_{\rm Q} = D_{\rm Q} \sum_{ \langle i, j \rangle }
[ ( \tilde{\tau}_i^Z \tilde{\tau}_j^Z + \tilde{\tau}_i^X \tilde{\tau}_j^X )
+ ( \tilde{\tau}_i^Y \tilde{\bsigma}_i ) \cdot ( \tilde{\tau}_j^Y \tilde{\bsigma}_j ) ], \\
& \tilde{H}_{\rm M} = D_{\rm M} \sum_{ \langle i, j \rangle }
[ \tilde{\bsigma}_i \cdot \tilde{\bsigma}_j
+ ( \tilde{\tau}_i^Z \tilde{\bsigma}_i ) \cdot ( \tilde{\tau}_j^Z \tilde{\bsigma}_j )
\nonumber \\
&~~~~~~~~~~~~~~~~~~~~~~~~
+ ( \tilde{\tau}_i^X \tilde{\bsigma}_i ) \cdot ( \tilde{\tau}_j^X \tilde{\bsigma}_j )
+ \tilde{\tau}_i^Y \tilde{\tau}_j^Y ],
\end{align}
where the pseudospin representations
$\tilde{\bsigma} = (\tilde{\sigma}^X, \tilde{\sigma}^Y, \tilde{\sigma}^Z) = 2 \bsigma$ and
$\tilde{\btau} = (\tilde{\tau}^X, \tilde{\tau}^Y, \tilde{\tau}^Z) = 2 \btau$ are defined in
Appendices~D and E.
For $\tilde{H_Z}$ in the three typical field directions,
\begin{align}
\tilde{H}_{\rm Z} = \left\{
\begin{array}{l}
\ds{ B \sum_i \left[ \frac{ 1 + 3 \lambda }{2} \tilde{\sigma}_i^Z
+ \frac{ 4 - 3 \lambda }{4} \tilde{\eta}_i^Z \right] }~~~
( \bH \parallel [001] \parallel Z ), \\
\ds{ B \sum_i \left[ \frac{ 1 - 2 \lambda }{2} \tilde{\sigma}_i^Z + \frac{ 2
+ \lambda }{2} \tilde{\eta}_i^Z
+ \frac{ 5 \lambda }{ 4 \sqrt{2} } ( \tilde{\sigma}_i^X + \tilde{\eta'}_i^X ) \right] } \\
~~~~~~~~~~~~~~~~~~~~~~~~~~~~~~~~~~~~~~~~~~~~~~~~~~( \bH \parallel [111] \parallel Z ), \\
\ds{ B \sum_i \frac{1}{16} \left[ ( 8 - 6 \lambda ) \tilde{\sigma}_i^Z
+ ( 16 + 3 \lambda ) \tilde{\eta}_i^Z - 15 \sqrt{3} \lambda \tilde{\zeta}_i^Z \right] } \\
~~~~~~~~~~~~~~~~~~~~~~~~~~~~~~~~~~~~~~~~~~~~~~~~~~( \bH \parallel [110] \parallel Z ).
\end{array}
\right.
\label{eqn:HZ3}
\end{align}
Here, $\tilde{\eta}^Z = \tilde{\tau}^Z \tilde{\sigma}^Z$,
$\tilde{\eta'}^X = \tilde{\tau}^Z \tilde{\sigma^X}$, and
$\tilde{\zeta}^Z = \tilde{\tau}^X \tilde{\sigma}^Z$
are new representations introduced to the octupole operators.
Note that $\tilde{\sigma}^X + \tilde{\eta'}^X$ and $\tilde{\zeta}^Z$ have off-diagonal matrix
elements of $\tilde{H}_Z$, as shown in Eqs.~(\ref{eqn:HZ110}) and (\ref{eqn:HZ111}).
\par

The primary order parameter is of the quadrupole type for $D_{\rm Q} > D_{\rm M}$.
According to the mean-field theory,
\cite{Shiina97}
the low-field phase is represented by the quadrupole component $O_u$ for $\bH \parallel [001]$,
$O_{yz} + O_{zx} + O_{xy}$ for $\bH \parallel [111]$, and the mixed $O_u$ and $O_{xy}$ for
$\bH \parallel [110]$.
In the $(XYZ)$ reference frame, all three quadrupole components have only diagonal matrix
elements of $\tilde{H}_Z$.
The resonance transition rate in Eq.~(\ref{eqn:Wn1}) is not affected by the quadrupole order;
the only change is a resonance energy shift.
The octupole effect causes the increase in the transition temperature $T_{\rm Q}$ with increasing
$H$, which can be derived from the GL expansion of the free energy in Appendix~F.
\par

\subsection{Effect of a staggered moment of antiferroquadrupole order on resonance transition
rate}
For the high-field phase in the highly symmetric interaction model for $D_Q = D_M$, a stable
quadrupole component is different from that of the low-field phase.
\cite{Shiina97}
One example is the quadrupole order for $\bH \parallel [110]$ associated with
$O_{yz} + O_{zx}$ that has off-diagonal matrix elements within the two lowest-lying states.
When this component is stabilized against the mixed $O_u \oplus O_{xy}$ component in the low
$H$, as mentioned in Sect.~3.3, it is necessary that the $\Gamma_5$-type
($O_{yz}, O_{zx}, O_{xy}$) intersite quadrupole--quadrupole coupling is slightly larger than the
$\Gamma_3$ type ($O_u$ and $O_v$).
Indeed, the linear combination of the $\Gamma_5$-type components is considered a prime
candidate for the quadrupole order parameter realized in CeB$_6$.
\cite{Shiina97,Sakai97,Shiina98,Matsumura12,Portnichenko20}
In the ($XYZ$) reference frame, $O_{yz} + O_{zx}$ corresponds to $O_{ZX}$ related to the
pseudospin representation:
$2 \tau^Y \sigma^Y = O_{ZX} / ( 2 \sqrt{3} )$.
For $\bH \parallel [110]$ in Table~I, $O_{ZX}$ has only matrix elements
between one pair of $| \psi_1 \rangle$ and $| \psi_2 \rangle$ and between another pair of
$| \psi_3 \rangle$ and $| \psi_4 \rangle$.
In the mean field theory, we consider a staggered moment $\mu_{\rm af}$ for the quadrupole
component $2 \tau^Y \sigma^Y$ in the local Hamiltonian
\begin{align}
H_{\rm local}^{(\pm)} = - B
\left(
\begin{array}{cc}
| E_1 | & \pm \bar{\mu} \\
\pm \bar{\mu} & | E_2 |
\end{array}
\right)~~\left( \bar{\mu} \equiv \frac{ D_Q  }{B} \mu_{\rm af} \right),
\end{align}
for the lowest-lying states $| \psi_1 \rangle$ and $| \psi_2 \rangle$.
Here, the signs $\pm$ of $\bar{\mu}$ are chosen for the two sublattice sites.
By diagonalizing $H_{\rm local}^{(\pm)}$, the eigenenergies and eigenfunctions are obtained as
\begin{align}
& \frac{ E_{\rm g} }{B} = -\frac{F_\lambda}{16} - \bar{E}_\mu~~~~~~
| \psi_{\rm g} \rangle = \cos \theta_\mu | \psi_1 \rangle \pm \sin \theta_\mu | \psi_2 \rangle,
\nonumber \\
& \frac{ E_{{\rm e}} }{B} = -\frac{F_\lambda}{16} + \bar{E}_\mu~~~~~~
| \psi_{{\rm e}} \rangle = \mp \sin \theta_\mu | \psi_1 \rangle + \cos \theta_\mu | \psi_2 \rangle,
\end{align}
for the ground and first excited states, respectively, where
\begin{align}
\bar{E}_\mu = \sqrt{ \left( \frac{ 4 - 3 \lambda }{8} \right)^2 + \bar{\mu}^2 },~~
\bar{E}_\mu \cos 2 \theta_\mu = \frac{ 4 - 3 \lambda }{8},~~
\bar{E}_\mu \sin 2 \theta_\mu = \bar{\mu}.
\end{align}
As shown in Sect.~3.2, we can calculate the rate $W_{{\rm e}, {\rm g}}^{[110]} ( \varphi )$ of the
acoustically driven resonance transition between $| \psi_{\rm g} \rangle$ and
$| \psi_{{\rm e}} \rangle$, and it has the same form as Eq.~(\ref{eqn:Wn1}).
The effect of the staggered moment $\mu_{\rm af}$ can be evaluated on the basis of the deviations
from the values of $A_{21}^{[110]}$, $B_{21}^{[110]}$, and $C_{21}^{[110]}$ for $\mu_{\rm af} = 0$
in Eq.~(\ref{eqn:ABC110}).
In particular, the amplitude $C_{{\rm e}, {\rm g}}^{[110]}$ of the $\cos 4 \varphi$ term in
$W_{{\rm e}, {\rm g}}^{[110]}$ is given by the simplest form
\begin{align}
\frac{ C_{{\rm e}, {\rm g}}^{[110]} }{ C_{21}^{[110]} }
= 1 + \frac{ 6 ( 4 - 3 \lambda )^2 }{F_\lambda^2}  \frac{ g_5^2 }{ g_3^2 } \sin^2 2 \theta_\mu.
\end{align}
For $\bar{\mu} \simeq 0$, it is reduced to
\begin{align}
\frac{ C_{{\rm e}, {\rm g}}^{[110]} }{ C_{21}^{[110]} }
= 1 + \frac{3}{2}
\frac{1}{ \left( 1 + \ds{ \frac{ 3 \lambda }{16} } \right)^2
+ \left( \ds{ \frac{ 15 \sqrt{3} \lambda }{16} } \right)^2 } \left( \frac{ g_5 }{ g_3 } \right)^2 \bar{\mu}^2.
\end{align}
This indicates that the amplitude of the $\cos 4 \varphi$ term in Eq.~(\ref{eqn:Wn1}) for the
two lowest-lying levels is increased by the induced quadrupole moment $\mu_{\rm af}$ in the
$O_{yz} + O_{zx}$ phase for $\bH \parallel [110]$.
Note that the transition rates between the $\Gamma_8$ ground and other excited
states are not changed for $\bH \parallel [110]$.
\par

\subsection{Photon-assisted single-phonon resonance transition}
\begin{figure}
\begin{center}
\includegraphics[width=7cm,clip]{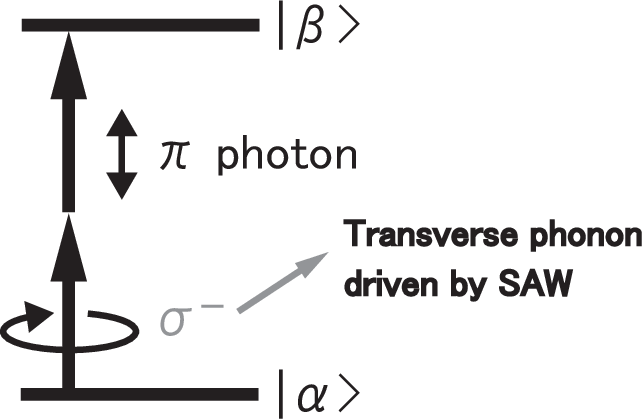}
\end{center}
\caption{
Analogy of photon-assisted single-phonon transition with the two-photon transition in which
left-hand circularly polarized ($\sigma^-$) and linearly polarized ($\pi$) photons are absorbed
between the two electronic states $| \alpha \rangle$ and $| \beta \rangle$.
The transverse phonon behaves like the $\sigma^-$ photon here.
The phonon (photon) absorption transition is of a nonmagnetic quadrupole (magnetic dipole)
type.
}
\label{fig:3}
\end{figure}
In comparison with the conventional ESR resonance, there are some limitations for increasing the
frequency of a strain field induced by SAW of gigahertz order.
Usually, an oscillating field of more than ten gigahertz is required to realize the spin resonance
transition in the presence of a strong magnetic field.
Indeed, this is also the case for CeB$_6$ where $B \sim 1$~T (order of frequency is ten gigahertz) is the typical energy scale for the resonance condition.
\par

A possible way of achieving acoustically driven spin resonance is to combine a strain field for the
transverse coupling with a linearly polarized microwave for the longitudinal coupling between
two levels (see Fig.~\ref{fig:3}).
The necessary resonance condition is given by
$\omega_A + \omega_L = \varepsilon_0 / \hbar$,
where $\omega_A$ and $\omega_L$ are the frequencies of the strain field and microwave,
respectively, and $\varepsilon_0$ is the energy of two-level splitting.
This is analogous to bichromatic driving for ESR using orthogonal electromagnetic wave fields.
\cite{Gromov00,Kalin04,Gyorgy22}
In our case, the oscillating photon magnetic field is parallel to the static magnetic field.
Even if $\omega_L = \varepsilon_0 / \hbar$ is satisfied, this $\pi$ photon absorption process cannot be the only source of the resonance transition.
The resonance requires a phonon absorption simultaneously with the $\pi$ photon, which is
strong evidence of transverse phonon coupling.
\par

The two-level Hamiltonian for the single-phonon transition combined with an oscillating $\pi$
photon field is written as
\begin{align}
H(t) = \frac{1}{2} \left(
\begin{array}{cc}
- \varepsilon_0 - \Delta \cos ( \omega_L t + \theta_L ) & A_T^* e^{ - i \omega_A t} \\
A_T e^{i \omega_A t} & \varepsilon_0 + \Delta \cos ( \omega_L t + \theta_L )
\label{eqn:Ht}
\end{array}
\right).
\end{align}
Here, $\Delta$ is the longitudinal coupling strength with a $\pi$ photon and $\theta_L$
represents a phase difference between the photon and phonon fields.
The rotating wave approximation has been applied to the transverse coupling with $A_T$,
which is only valid for describing single-phonon absorption.
The weak coupling $| A_T | \ll \hbar \omega_A$ is also assumed.
In the two-level system comprising $| \alpha \rangle$ and $| \beta \rangle$, $A_T$ corresponds
to an off-diagonal matrix element of the spin--strain interaction Hamiltonian
\begin{align}
\frac{ A_T }{2} = \langle \beta | \tilde{H}_\varepsilon | \alpha \rangle.
\end{align}
When we choose the two states $| \alpha \rangle = | \psi_1 \rangle$ and
$| \beta \rangle = | \psi_2 \rangle$ for $\bH \parallel [110]$ in Table~I, we have
\begin{align}
A_T = - i 2 \sqrt{3} \left[ \pm g_5 \varepsilon_{x'z}'' \cos \left( \varphi - \frac{ \pi }{4} \right)
- g_3 \varepsilon_{x'x'} \cos 2 \varphi \right]
\label{eqn:ATH110}
\end{align}
for the SAWs propagating in the $\pm x'$ directions.
\par

In the rotating frame of the phonon field, the transformed Hamiltonian for Eq.~(\ref{eqn:Ht}) is given
by
\begin{align}
\tilde{H} (t) = \frac{1}{2} \left(
\begin{array}{cc}
- \varepsilon (t)  & A_T^* \\
A_T & \varepsilon (t)
\end{array}
\right),
\label{eqn:tildeH}
\end{align}
where $\varepsilon (t) = \varepsilon_{0 A} + \Delta \cos ( \omega t + \theta )$ with the detuning
$\varepsilon_{0 A} = \varepsilon_0 - \hbar \omega_A$ and the subscript $L$ is omitted for
$\omega_L$ and $\theta_L$.
For the periodically time-dependent Hamiltonian, the Schr\"{o}dinger equation
\begin{align}
i \hbar \frac{ \partial \psi ( t ) }{ \partial t } = \tilde{H} (t) \psi ( t )
\end{align}
can be solved using the Floquet theory.
\cite{Shirley65,Son09}
Substituting $\psi (t ) = e^{ - i q t } \phi (t)$ into this equation, we obtain
\begin{align}
\left[ \tilde{H} (t) - i \hbar \frac{ \partial }{ \partial t } \right] \phi (t) = q \phi (t)
\label{eqn:eqn-q}
\end{align}
to solve an eigenvalue problem for the quasienergy $q$.
By applying the Fourier expansion to $\tilde{H} (t)$ and $\phi (t)$ as
\begin{align}
\tilde{H} (t) = \sum_n H^{ [n] } e^{ i n \omega t },~~
\phi (t) = \sum_n \phi^{ [n] } e^{ i n \omega t },
\end{align}
the problem of solving Eq.~(\ref{eqn:eqn-q}) is transformed to the following
time-independent eigenvalue problem:
\begin{align}
\sum_m ( H^{ [ n - m ] } + n \hbar \omega \delta_{nm} ) \phi^{ [ m ] } = q \phi^{ [ n ] }.
\end{align}
This can be rewritten using the so-called Floquet Hamiltonian $H_F$ as
\begin{align}
\langle \alpha n | H_F | \beta m \rangle = H_{\alpha \beta}^{ [ n - m] }
+ n \hbar \omega \delta_{nm} \delta_{\alpha \beta}.
\label{eqn:HF}
\end{align}
Indeed, $H_F$ has an infinite-dimensional matrix form for the Floquet states
$| \alpha n \rangle = | \alpha \rangle \otimes | n \rangle$, where $\alpha$ denotes one of the two
levels and $n$ ($ = 0, \pm 1, \pm 2, \cdots $) represents the Fourier indices.
Thus, the eigenvalue problem is described by solving
$H_F | q_\gamma \rangle = q_\gamma | q_\gamma \rangle$ with the $\gamma$th eigenvalue
$q_\gamma$ and the corresponding eigenvector $| q_\gamma \rangle$.
In the following discussion, $\hbar = 1$ is set for brevity.
\par

For the effective Hamiltonian in Eq.~(\ref{eqn:tildeH}), the nonvanishing Fourier components are
$H^{ [ n - m ] }$ in Eq.~(\ref{eqn:HF}) for $n - m = 0, \pm 1$, and these three matrices are given by
\begin{align}
H^{ [0] } = \frac{1}{2}
\left(
\begin{array}{cc}
- \varepsilon_{0 A} & A_T^* \\
A_T & \varepsilon_{0 A}
\end{array}
\right),~~
H^{ [ \pm 1 ] } = \frac{1}{4}
\left(
\begin{array}{cc}
- \Delta e^{ \pm i \theta } & 0 \\
0 & \Delta e^{ \pm i \theta }
\end{array}
\right).
\end{align}
With the $2 \times 2$ block matrices, the matrix form of the Floquet Hamiltonian is written as
\begin{align}
H_F =
\left(
\begin{array}{ccccccc}
\ddots & ~ & ~ & \vdots & ~ & ~ & ~ \\
~ & H_{-2}^{[0]} & H^{[-1]} & {\bf 0} & {\bf 0} & {\bf 0} & ~ \\
~ & H^{[1]} & H_{-1}^{[0]} & H^{[-1]} & {\bf 0} & {\bf 0} & ~ \\
\cdots & {\bf 0} & H^{[1]} & H_0^{[0]} & H^{[-1]} & {\bf 0} & \cdots \\
~ & {\bf 0} & {\bf 0} & H^{[1]} & H_1^{[0]} & H^{[-1]} & ~ \\
~ & {\bf 0} & {\bf 0} & {\bf 0} & H^{[1]} & H_2^{[0]} & ~ \\
~ & ~ & ~ & \vdots & ~ & ~ & \ddots
\end{array}
\right),
\label{eqn:HFblock}
\end{align}
where $H_n^{ [0] } \equiv H^{ [0] } + n \omega I$ ($I$: identity matrix) and ${\bf 0}$ represents the
zero matrix.
To analyze the Floquet matrix, we use the eigenstates of $H_F$ for $A_T = 0$,
\begin{align}
& | \alpha n \rangle_L = \sum_{ k = - \infty }^\infty e^{ i ( k - n ) \theta }
J_{k - n} \left( \frac{ \Delta }{ 2 \omega } \right) | \alpha k \rangle,
\nonumber \\
& | \beta m \rangle_L = \sum_{ k = - \infty }^\infty e^{ i ( k - m ) \theta }
J_{k - m} \left( - \frac{ \Delta }{ 2 \omega } \right) | \beta k \rangle,
\end{align}
where $J_k$ is the $k$th Bessel function of the first kind.
In the present theory, we consider the resonance transition probability under the condition
$- \varepsilon_{0 A} / 2 \simeq \varepsilon_{0 A} / 2 - n \omega$, indicating that the nearly
degenerate Floquet states, for instance, $| \alpha 0 \rangle_L$ and $| \beta, - n \rangle_L$,
are coupled through an off-diagonal matrix element including $J_{-n} ( \Delta / \omega )$.
On the basis of the Van Vleck perturbation theory, the infinite-dimensional matrix of $H_F$ is
reduced to the following $2 \times 2$ matrix form in the subspace of $| \alpha 0 \rangle_L$ and
$| \beta, - n \rangle_L$:
\begin{align}
\tilde{H}_F = \left(
\begin{array}{cc}
\ds{ - \frac{ \varepsilon_{0 A} }{2} } + \delta_n & v_{ - n }^* \\
v_{ - n } & \ds{ \frac{ \varepsilon_{0 A} }{2} } - \delta_n - n \omega
\end{array}
\right),
\end{align}
where the off-diagonal matrix element $v_{ - n }$ and the energy shift $\delta_n$ can be
expanded by choosing $A_T / 2$ as a perturbation parameter.
Their leading terms are given as
\cite{Son09}
\begin{align}
v_{ - n } = \frac{ A_T }{2} e^{ - i n \theta } J_{ - n } \left( \frac{ \Delta }{ \omega } \right),~~
\delta_n = - \sum_{ {}_{ k \ne - n }^{ k  = - \infty } }^\infty
\frac{ | v_k |^2 }{ \varepsilon_{0 A} + k \omega }.
\end{align}
After diagonalizing $\tilde{H}_F$, we obtain the eigenvalues
\begin{align}
q_\pm = - \frac{ n \omega }{2} \pm + \tilde{q}_n,
\end{align}  
where
\begin{align}
\tilde{q}_n = \sqrt{ \frac{ ( n \omega - \varepsilon_{0A} + 2 \delta_n )^2 }{4} + | v_{ - n } |^2 }.
\label{eqn:qn}
\end{align}
Equation~(\ref{eqn:qn}) leads to the $n$ $\pi$-photon time-averaged transition probability from
$| \alpha \rangle$ to $| \beta \rangle$,
\cite{Son09}
\begin{align}
\bar{P}_{\alpha \rightarrow \beta}^{ (n) } =
\frac{1}{2} \cdot \frac{ | v_{ - n } |^2 }
{ \ds{ \frac{ ( n \omega - \varepsilon_{0A} + 2 \delta_n )^2 }{4} } + | v_{ - n } |^2 }.
\end{align}
For the single $\pi$ photon absorption,
\begin{align}
\bar{P}_{\alpha \rightarrow \beta}^{(1)} =
\frac{1}{2} \frac{ | A_T J_1 |^2 }{ | A_T J_1 |^2 + ( \omega - \varepsilon_{0A} + 2 \delta_1 )^2 }.
\end{align}
Here, $J_{-1} ( x ) = - J_1 (x)$ is used and $J_1 ( \Delta / \omega )$ is denoted by $J_1$.
For the weak coupling limit ($| A_T |, \Delta \ll \omega$), $\delta_1$ and $J_1$ are approximated
as
\begin{align}
\delta_1 \simeq - \frac{ | A_T |^2 }{ 4 \varepsilon_{0A} },~~
J_1 \simeq \frac{ \Delta }{ 2 \omega }.
\end{align}
Except for $A_T = 0$, the transition probability at $\omega = \varepsilon_{0A}$ is obtained as
\begin{align}
\bar{P}_{\alpha \rightarrow \beta}^{(1)} =
\frac{1}{2} \frac{1}{ 1 + ( | A_T | / \Delta )^2 }~~( \varepsilon_0 = \omega + \omega_{A} ).
\end{align}
As given by Eq.~(\ref{eqn:ATH110}), the $\varphi$ dependence of $| A_T |^2$ can be evaluated
from the transition probabilities at $\varepsilon_0 = \omega + \omega_{A}$ measured for various
propagation directions of SAW.
Even for large excitation energy gaps $\varepsilon_0$ under strong magnetic fields, the MAR
is available with the help of optical absorption.
The detailed description of this method is beyond the scope of this study.

\section{Conclusion}
Motivated by the recent progress in acoustic measurements using SAWs, we presented a new
idea of MAR to probe multipole degrees of freedom in a crystal field quartet such as an $O_h$
$\Gamma_8$ representation.
In particular, we focused on an octupole effect on the Zeeman splitting of the quartet, which was
explicitly expressed by the anisotropy parameter $\lambda$ here.
We demonstrated how to evaluate $\lambda$ as well as the ratio of the quadrupole--strain
coupling strengths $g_5 / g_3$ from the dependence of the resonance transition rate on the SAW
propagation direction represented by $\varphi$.
The amplitude of the $\cos 4 \varphi$ term is a key in the quantitative evaluation.
It also provides useful information about the identification of a primary order parameter for
multipole--multipole interaction systems.
As a simple case, we demonstrated how to confirm an induced $\Gamma_5$-type
quadrupole-ordered moment by measuring the amplitude of the $\cos 4 \varphi$ term in the MAR.
Finally, we proposed a method of photon-assisted single-phonon resonance formulated on the
basis of the Floquet theory.
This method of using a $\pi$ photon field is particularly useful for the detection of the
acoustically driven phonon transition for rather large excitation energy gaps under strong 
magnetic fields.
\par

The significant role of octupole degrees of freedom has been studied extensively for the
antiferroquadrupole order phase II in CeB$_6$.
In particular, the nuclear magnetic resonance line splitting measured at the B sites was
successfully explained by considering not only dipole but also octupole moments of a
$\Gamma_8$ ground state induced by a magnetic field in the quadrupole order phase.
\cite{Takigawa83,Sakai97,Shiina98}
The direct evidence of field-induced octupoles was provided by resonant X-ray diffraction
measurements
\cite{Matsumura12}
and inelastic neutron scattering under the rotation of the field direction
\cite{Portnichenko20}.
On the other hand, some open questions about the local $\Gamma_8$ scenario have been
raised by the observation of ESR signals in a high-field phase
\cite{Semeno16,Schlottmann18,Semeno21}
and the recent experiments using B-site nuclear quadrupole resonance in a very low field.
\cite{Mito23}
So far, no acoustic measurement has been carried out as a microscopic probe of the multipoles.
The MAR proposed here can provide direct evidence of whether the local $\Gamma_8$ quartet
is the origin of the observed multipole features.
For the photon-assisted MAR in CeB$_6$, the frequency of the typical energy scale
($10$--$100$~GHz) is accessible by conventional methods using microwaves.
\par

As mentioned in Sect.~1, the V$_{\rm Si}$ quartet is a good candidate for the investigation of
the fine structure of the spin-3/2 multipole origin.
\cite{Soltamov19,Simin16,Tarasenko18}
In particular, a characteristic of the spin quadrupole was actually detected as anisotropic
acoustically driven spin resonances at room temperature.
\cite{Hernandez-Minguez20}
The $V_{\rm Si}$ quartet in SiC has the $C_{3v}$ site symmetry, which makes the $V_{\rm Si}$
spin--strain interaction model more complicated owing to the symmetry lowering from $O_h$.
\cite{Koga24,Udvarhelyi18a,Udvarhelyi18b}
The dipole and octupole coupling strengths of the Zeeman terms (anisotropic $g$-factors) are different
between the two components parallel to and orthogonal to a typical crystallographic axis.
\cite{Simin16,Tarasenko18}
By considering a complete set of anisotropies associated with the crystal field symmetry, we will
develop our method of MAR to reveal a characteristic of the octupole hidden in various defect
multiplets.

\bigskip
{\footnotesize
{\bf Acknowledgment}~~This work was supported by JSPS KAKENHI Grant Number 21K03466.
}

\appendix
\section{Multipole operators for $J = 3/2$}
Here, the irreducible tensor operator representations are listed for the dipole
$\{ J_x, J_y, J_z \}$, quadrupole $\{ O_u, O_v, O_{zx}, O_{xy}, O_{yz} \}$, and octupole
$\{ T_x^\alpha, T_y^\alpha, T_z^\alpha, T_x^\beta, T_y^\beta, T_z^\beta, T_{xyz} \}$ in the
$O_h$ point group.
\cite{Shiina97}
The corresponding $4 \times 4$ matrices are also presented as follows.
For the $\Gamma_4$ dipole components,
\begin{align}
& J_x = \frac{1}{ \sqrt{2} } ( - J_1^{(1)} + J_{-1}^{(1)} )
\nonumber \\
&~~~~
= \frac{1}{2}
\left(
\begin{array}{cccc}
0 & \sqrt{3} & 0 & 0 \\
\sqrt{3} & 0 & 2 & 0 \\
0 & 2 & 0 & \sqrt{3} \\
0 & 0 & \sqrt{3} & 0
\end{array}
\right),
\nonumber \\
& J_y = \frac{i}{ \sqrt{2} } ( J_1^{(1)} + J_{-1}^{(1)} )
\nonumber \\
&~~~~
= \frac{1}{2}
\left(
\begin{array}{cccc}
0 & -i \sqrt{3} & 0 & 0 \\
i \sqrt{3} & 0 & -i 2 & 0 \\
0 & i 2 & 0 & -i \sqrt{3} \\
0 & 0 & i \sqrt{3} & 0
\end{array}
\right),
\nonumber \\
& J_z = J_0^{(1)}
\nonumber \\
&~~~~
= \frac{1}{2}
\left(
\begin{array}{cccc}
3 & 0 & 0 & 0 \\
0 & 1 & 0 & 0 \\
0 & 0 & -1 & 0 \\
0 & 0 & 0 & -3
\end{array}
\right).
\label{eqn:Jmu}
\end{align}
For the $\Gamma_3$ and $\Gamma_5$ quadrupole components,
\begin{align}
& O_u = \frac{1}{ \sqrt{3} } ( 2 J_z^2 - J_x^2 - J_y^2 ) = \frac{2}{ \sqrt{3} } J_0^{(2)}
\nonumber \\
&~~~~~~
= \sqrt{3}
\left(
\begin{array}{cccc}
1 & 0 & 0 & 0 \\
0 & -1 & 0 & 0 \\
0 & 0 & -1 & 0 \\
0 & 0 & 0 & 1
\end{array}
\right),
\nonumber \\
& O_v = J_x^2 - J_y^2 = \sqrt{ \frac{2}{3} } ( J_2^{(2)} + J_{-2}^{(2)} )
\nonumber \\
&~~~~~~
= \sqrt{3}
\left(
\begin{array}{cccc}
0 & 0 & 1 & 0 \\
0 & 0 & 0 & 1 \\
1 & 0 & 0 & 0 \\
0 & 1 & 0 & 0
\end{array}
\right),
\end{align}
and
\begin{align}
& O_{zx} = J_z J_x  + J_x J_z =  \sqrt{ \frac{2}{3} } ( - J_1^{(2)} + J_{-1}^{(2)} )
\nonumber \\
&~~~~~~
= \sqrt{3}
\left(
\begin{array}{cccc}
0 & 1 & 0 & 0 \\
1 & 0 & 0 & 0 \\
0 & 0 & 0 & -1 \\
0 & 0 & -1 & 0
\end{array}
\right),
\nonumber \\
& O_{xy} = J_x J_y  + J_y J_x =  i \sqrt{ \frac{2}{3} } ( - J_2^{(2)} + J_{-2}^{(2)} )
\nonumber \\
&~~~~~~
= \sqrt{3}
\left(
\begin{array}{cccc}
0 & 0 & -i & 0 \\
0 & 0 & 0 & -i \\
i & 0 & 0 & 0 \\
0 & i & 0 & 0
\end{array}
\right),
\nonumber \\
& O_{yz} = J_y J_z  + J_z J_y = i  \sqrt{ \frac{2}{3} } ( J_1^{(2)} + J_{-1}^{(2)} )
\nonumber \\
&~~~~~~
= \sqrt{3}
\left(
\begin{array}{cccc}
0 & -i & 0 & 0 \\
i & 0 & 0 & 0 \\
0 & 0 & 0 & i \\
0 & 0 & -i & 0
\end{array}
\right),
\end{align}
respectively.
The $\Gamma_4$ octupole components are given by
\begin{align}
& T_x^\alpha = \frac{1}{4} [ \sqrt{5} ( - J_3^{(3)} + J_{-3}^{(3)} )
- \sqrt{3} ( - J_1^{(3)} + J_{-1}^{(3)} ) ]
\nonumber \\
&~~~~~~
= \frac{3}{8}
\left(
\begin{array}{cccc}
0 & -\sqrt{3} & 0 & 5 \\
-\sqrt{3} & 0 & 3 & 0 \\
0 & 3 & 0 & -\sqrt{3} \\
5 & 0 & -\sqrt{3} & 0
\end{array}
\right),
\nonumber \\
& T_y^\alpha = - \frac{i}{4} [ \sqrt{5} ( J_3^{(3)} + J_{-3}^{(3)} )
+ \sqrt{3} ( J_1^{(3)} + J_{-1}^{(3)} ) ]
\nonumber \\
&~~~~~~
= \frac{3}{8}
\left(
\begin{array}{cccc}
0 & i \sqrt{3} & 0 & i 5 \\
-i \sqrt{3} & 0 & -i 3 & 0 \\
0 & i 3 & 0 & i \sqrt{3} \\
-i 5 & 0 & -i \sqrt{3} & 0
\end{array}
\right),
\nonumber \\
& T_z^\alpha = J_0^{(3)}
\nonumber \\
&~~~~
= \frac{3}{4}
\left(
\begin{array}{cccc}
1 & 0 & 0 & 0 \\
0 & -3 & 0 & 0 \\
0 & 0 & 3 & 0 \\
0 & 0 & 0 & -1
\end{array}
\right).
\label{eqn:Talpha}
\end{align}
For $\Gamma_5$,
\begin{align}
& T_x^\beta = - \frac{1}{4} [ \sqrt{3} ( - J_3^{(3)} + J_{-3}^{(3)} )
+ \sqrt{5} ( - J_1^{(3)} + J_{-1}^{(3)} ) ]
\nonumber \\
&~~~~~~
= - \frac{ 3 \sqrt{5} }{8}
\left(
\begin{array}{cccc}
0 & 1 & 0 & \sqrt{3} \\
1 & 0 & - \sqrt{3} & 0 \\
0 & - \sqrt{3} & 0 & 1 \\
\sqrt{3} & 0 & 1 & 0
\end{array}
\right),
\nonumber \\
& T_y^\beta = \frac{i}{4} [ - \sqrt{3} ( J_3^{(3)} + J_{-3}^{(3)} )
+ \sqrt{5} ( J_1^{(3)} + J_{-1}^{(3)} ) ]
\nonumber \\
&~~~~~~
= \frac{ 3 \sqrt{5} }{8}
\left(
\begin{array}{cccc}
0 & -i & 0 & i \sqrt{3} \\
i & 0 & i \sqrt{3} & 0 \\
0 & -i\sqrt{3} & 0 & -i \\
-i \sqrt{3} & 0 & i & 0
\end{array}
\right),
\nonumber \\
& T_z^\beta = \frac{1}{ \sqrt{2} } ( J_2^{(3)} + J_{-2}^{(3)} )
\nonumber \\
&~~~~
= \frac{ 3 \sqrt{5} }{4}
\left(
\begin{array}{cccc}
0 & 0 & 1 & 0 \\
0 & 0 & 0 & -1 \\
1 & 0 & 0 & 0 \\
0 & -1 & 0 & 0
\end{array}
\right).
\end{align}
Lastly, the $\Gamma_2$ representation has a single component:
\begin{align}
& T_{xyz} = \frac{i}{ \sqrt{2} } ( - J_2^{(3)} + J_{-2}^{(3)} )
\nonumber \\
&~~~~
= \frac{ 3 \sqrt{5} }{4}
\left(
\begin{array}{cccc}
0 & 0 & -i & 0 \\
0 & 0 & 0 & i \\
i & 0 & 0 & 0 \\
0 & -i & 0 & 0
\end{array}
\right).
\end{align}

\section{Unitary transformation for field direction as a quantization axis}
The unitary transformation introduced in Sect~2.2 is expressed in the $4 \times 4$ form,
\begin{align}
U = \left(
\begin{array}{cccc}
\chi^{-3} c c_{a+} & \chi^{-3} c \tilde{s} & \chi^{-3} s \tilde{s} & \chi^{-3} s c_{a-} \\
\chi^{-1} c \tilde{s} & \chi^{-1}  c c_{b-} & - \chi^{-1}  s c_{b+} & - \chi^{-1} s \tilde{s} \\
\chi s \tilde{s} & - \chi s c_{b+} & - \chi c c_{b-} & \chi c \tilde{s} \\
\chi^3 s c_{a-} & - \chi^3 s \tilde{s} & \chi^3 c \tilde{s} & - \chi^3 c c_{a+}
\end{array}
\right).
\label{eqn:US}
\end{align}
Here, each matrix element comprises the following terms:
\begin{align}
& \chi = e^{i \phi / 2},~~c = \cos \frac{ \theta }{2},~~s = \sin \frac{ \theta }{2},~~
\tilde{s} = \frac{ \sqrt{3} }{2} \sin \theta,
\nonumber \\
& c_{a \pm} = \frac{ 1 \pm \cos \theta }{2},~~c_{b \pm} = \frac{ 1 \pm 3 \cos \theta }{2}.
\end{align}
The Zeeman term is anisotropic with respect to the field direction owing to the octupole
components.
For the transformed Zeeman term in Eq.~(\ref{eqn:UTB}), the coefficients
$f_i$ ($i = 1, 2, \cdots, 7$) are listed as
\begin{align}
& f_1 = \frac{B}{8} \left[ ( 3 - 30 \cos^2 \theta + 35 \cos^4 \theta ) + 5 \sin^4 \theta \cos 4 \phi \right],
\nonumber \\
& f_2 = \frac{ \sqrt{15} }{8} B \sin^2 \theta \left[ ( -1 + 7 \cos^2 \theta )
+ ( 1 + \cos^2 \theta ) \cos 4 \phi \right],
\nonumber \\
& f_3 = - \frac{ \sqrt{15} }{4} B \sin^2 \theta \cos \theta \sin 4 \phi,
\nonumber \\
& f_4 = \frac{ 5 \sqrt{6} }{16} B \sin \theta \cos \theta
\left[ ( 3 - 7 \cos^2 \theta ) + \sin^2 \theta \cos 4 \phi \right],
\nonumber \\
& f_5 = \frac{ \sqrt{10} }{16} B \sin \theta \cos \theta
\left[ 7 \sin^2 \theta - ( 3 + \cos^2 \theta ) \cos 4 \phi \right],
\nonumber \\
& f_6 = - \frac{ 5 \sqrt{6} }{16} B \sin^3 \theta \sin 4 \phi,
\nonumber \\
& f_7 = \frac{ \sqrt{10} }{16} B \sin \theta ( 1 + 3 \cos^2 \theta ) \sin 4 \phi.
\label{eqn:TBf}
\end{align}

\section{Unitary transformed quadrupole operators and strain-dependent coefficients}
In Sect.~2.2, the unitary transformation has been introduced into the dipole and octupole terms in
$H_Z$, as shown in Eqs.~(\ref{eqn:UJB}) and (\ref{eqn:UTB}), respectively.
For the quadrupole operators in Eq.~(\ref{eqn:Hep}), the unitary transformed $O_k$
($k = u, v, zx, xy, yz$) are written as 
\begin{align}
& U^\dagger O_u U = - \frac{ 1 - 3 \cos^2 \theta }{2} O_U + \frac{ \sqrt{3} }{2} \sin^2 \theta O_V
\nonumber \\
&~~~~~~~~~~~~~~
+ \sqrt{3} \sin \theta \cos \theta O_{ZX},
\nonumber \\
& U^\dagger O_v U = \left( \frac{ \sqrt{3} }{2} \sin^2 \theta O_U
+ \frac{ 1 + \cos^2 \theta }{2} O_V - \sin \theta \cos \theta O_{ZX} \right)
\nonumber \\
&~~~~~~~~~~~~~~~~~~~~
\times  \cos 2 \phi
\nonumber \\
&~~~~~~~~~~~~~~~
+ ( - \cos \theta O_{XY} + \sin \theta O_{YZ} )  \sin 2 \phi,
\nonumber \\
& U^\dagger O_{zx} U = ( \sqrt{3} \sin \theta \cos \theta O_U - \sin \theta \cos \theta O_V
- \cos 2 \theta O_{ZX} )
\nonumber \\
&~~~~~~~~~~~~~~~~~~~~
\times \cos \phi
\nonumber \\
&~~~~~~~~~~~~~~
+ ( \sin \theta O_{XY} + \cos \theta O_{YZ} ) \sin \phi,
\nonumber \\
& U^\dagger O_{xy} U = \left( \frac{ \sqrt{3} }{2} \sin^2 \theta O_U
+ \frac{ 1 + \cos^2 \theta }{2} O_V - \sin \theta \cos \theta O_{ZX} \right)
\nonumber \\
&~~~~~~~~~~~~~~~~~~~~
\times \sin 2 \phi
\nonumber \\
&~~~~~~~~~~~~~~
+ ( \cos \theta O_{XY} - \sin \theta O_{YZ} ) \cos 2 \phi,
\nonumber \\
& U^\dagger O_{yz} U = ( \sqrt{3} \sin \theta \cos \theta O_U - \sin \theta \cos \theta O_V
- \cos 2 \theta O_{ZX} )
\nonumber \\
&~~~~~~~~~~~~~~~~~~~~
\times \sin \phi
\nonumber \\
&~~~~~~~~~~~~~~
+ ( - \sin \theta O_{XY} - \cos \theta O_{YZ} ) \cos \phi.
\label{eqn:OkOK}
\end{align}
Accordingly, the strain-dependent coupling coefficients $A_K$ have the following
field-direction dependences:
\begin{align}
& A_U = g_3 \left( -  \varepsilon_u \frac{1 - 3 \cos^2 \theta }{2} + \frac{ \sqrt{3} }{2}
\varepsilon_v \sin^2 \theta \cos 2 \phi \right)
\nonumber \\
&~~~~
+ g_5 \Bigg[ \frac{ \sqrt{3} }{2}  \varepsilon_{xy} \sin^2 \theta \sin 2 \phi
\nonumber \\
&~~~~~~~~~~
+ \sqrt{3} \sin \theta \cos \theta ( \varepsilon_{zx} \cos \phi + \varepsilon_{yz} \sin \phi ) \Bigg],
\nonumber \\
& A_V = g_3 \left( \frac{ \sqrt{3} }{2} \varepsilon_u \sin^2 \theta  + \varepsilon_v
\frac{1 + \cos^2 \theta }{2} \cos 2 \phi \right)
\nonumber \\
&~~~~
+ g_5 \left[ \varepsilon_{xy} \frac{ 1 + \cos^2 \theta }{2} \sin 2 \phi
- \sin \theta \cos \theta ( \varepsilon_{zx} \cos \phi + \varepsilon_{yz} \sin \phi ) \right],
\nonumber \\
& A_{ZX} = g_3 \sin \theta \cos \theta ( \sqrt{3} \varepsilon_u - \varepsilon_v \cos 2 \phi )
\nonumber \\
&~~~~~~
+ g_5 \left[ - \varepsilon_{xy} \sin \theta \cos \theta \sin 2 \phi
- \cos 2 \theta ( \varepsilon_{zx} \cos \phi + \varepsilon_{yz} \sin \phi ) \right],
\nonumber \\
& A_{XY} = - g_3 \varepsilon_v \cos \theta \sin 2 \phi
\nonumber \\
&~~~~~~
+ g_5 \left[ \varepsilon_{xy} \cos \theta \cos 2 \phi
+ \sin \theta ( \varepsilon_{zx} \sin \phi - \varepsilon_{yz} \cos \phi ) \right],
\nonumber \\
& A_{YZ} = g_3 \varepsilon_v \sin \theta \sin 2 \phi
\nonumber \\
&~~~~~~
+ g_5 \left[ - \varepsilon_{xy} \sin \theta \cos 2 \phi
+ \cos \theta ( \varepsilon_{zx} \sin \phi - \varepsilon_{yz} \cos \phi ) \right].
\label{eqn:AK}
\end{align}

\section{Pseudospin-1/2 representation of the $\Gamma_8$ quartet and its relation to
multipole operators for $J = 3/2$}
Following Shiina {\it et al.},
\cite{Shiina97}
it is convenient to introduce two kinds of pseudospin-1/2 components representing spin
($\uparrow, \downarrow$) and orbital ($\oplus, \ominus$) degrees of freedom in the $\Gamma_8$
quartet.
We connect the pseudospin representations to the multipole operators constructed from the
$J = 3/2$ operators for the quartet.
This is different from the way of using the original $J = 5/2$ operators in the earlier study.
\cite{Shiina97}
\par

First, the $\Gamma_8$ quartet is denoted by
\begin{align}
&| \Gamma_{8,1} \rangle \rightarrow | 3/2 \rangle_{ J_{\Gamma_8} = 3/2 }
\rightarrow - | \oplus \downarrow \rangle,
\nonumber \\
& | \Gamma_{8,2} \rangle \rightarrow | 1/2 \rangle_{ J_{\Gamma_8} = 3/2 }
\rightarrow | \ominus \uparrow \rangle,
\nonumber \\
& | \Gamma_{8,3} \rangle \rightarrow | - 1/2 \rangle_{ J_{\Gamma_8} = 3/2 }
\rightarrow - | \ominus \downarrow \rangle,
\nonumber \\
& | \Gamma_{8,4} \rangle \rightarrow | - 3/2 \rangle_{ J_{\Gamma_8} = 3/2 }
\rightarrow | \oplus \uparrow \rangle.
\label{eqn:G8}
\end{align}
The two components $\sigma (= \uparrow, \downarrow )$ are connected to each other by Pauli
spin matrices $\bsigma = ( \sigma^x, \sigma^y, \sigma^z )$, and $\tau (= \oplus, \ominus )$ are
connected by $\btau = ( \tau^x, \tau^y, \tau^z )$ in the same manner.
Each spin component and a product of two components $\tau^\alpha$ and $\sigma^\beta$
($\alpha, \beta = x, y, z$) are related to the multipole operators constructed by $J = 3/2$ operators,
which can be categorized by the irreducible representations of the $O_h$ point group.
For $\Gamma_3$ and $\Gamma_5$ quadrupole components,
\begin{align}
\tau^z = \frac{1}{ 2 \sqrt{3} } O_u,~~\tau^x = \frac{1}{ 2 \sqrt{3} } O_v,
\end{align}
and
\begin{align}
2 \tau^y \sigma^x = - \frac{1}{ 2 \sqrt{3} } O_{yz},~~
2 \tau^y \sigma^y = - \frac{1}{ 2 \sqrt{3} } O_{zx},~~
2 \tau^y \sigma^z = - \frac{1}{ 2 \sqrt{3} } O_{xy},
\end{align}
respectively.
The linear combination of dipole and octupole components with the same representation
$\Gamma_4$ are given by
\begin{align}
\sigma^\mu = -\frac{1}{5} J_\mu - \frac{4}{15} T_\mu^\alpha~~(\mu = x, y, z),
\end{align}
and
\begin{align}
& 2 \tau^z \sigma^z = - \frac{2}{5} J_z + \frac{2}{15} T_z^\alpha,
\nonumber \\
& - \tau^z \sigma^x + \sqrt{3} \tau^x \sigma^x = - \frac{2}{5} J_x + \frac{2}{15} T_x^\alpha,
\nonumber \\
& - \tau^z \sigma^y - \sqrt{3} \tau^x \sigma^y = - \frac{2}{5} J_y + \frac{2}{15} T_y^\alpha.
\label{eqn:G4JT}
\end{align}
The other multipole components are $\Gamma_5$ and $\Gamma_2$ types written as
\begin{align}
& 2 \tau^x \sigma^z = - \frac{2}{ 3\sqrt{5} } T_z^\beta,
\nonumber \\
& {- \sqrt{3}} \tau^z \sigma^x - \tau^x \sigma^x = - \frac{2}{ 3\sqrt{5} } T_x^\beta,
\nonumber \\
& \sqrt{3} \tau^z \sigma^y - \tau^x \sigma^y = - \frac{2}{ 3\sqrt{5} } T_y^\beta,
\end{align}
and
\begin{align}
\tau^y = \frac{2}{ 3\sqrt{5} } T_{xyz},
\end{align}
respectively.
As listed above, our representations are similar to those in Ref.~3 for the relationships between
the $\tau \otimes \sigma$ representations and multipole components. 

\section{Multipole--multipole interaction model}
The spherical multipole--multipole interaction is invariant under the rotation of a magnetic field.
On the other hand, the Zeeman splitting is anisotropic for the $\Gamma_8$ quartet in
Eq.~(\ref{eqn:G8}) owing to the different octupole components in the Zeeman term, as shown
in Eq.~(\ref{eqn:HZ3}).
This is a crucial point for the multipole symmetry of an induced ordered moment depending on the
field direction.
\par

The highly spherical interaction is given by the following Hamiltonian:
\cite{Shiina97,Ohkawa83,Ohkawa85}
\begin{align}
& H_{\rm s} = D \sum_{\langle i, j \rangle} [ \bsigma_i \cdot \bsigma_j + \btau_i \cdot \btau_j
+ 4 ( \btau_i \cdot \btau_j ) ( \bsigma_i \cdot \bsigma_j ) ] + H_{\rm Z}.
\label{eqn:Hs}
\end{align}
Here, the first term consists of spin, orbital, and spin $+$ orbital interactions between the
nearest-neighbor $i$ and $j$ sites with coupling constant $D$ ($>0$).
This interaction is SU(4) symmetric with respect to the $4 \times 4$ pseudospin space
$\tau \otimes \sigma$. 
The last term $H_{\rm Z}$ represents the Zeeman splitting of the quartet denoted by
$| \tau, \sigma \rangle$,
\begin{align}
H_{\rm Z} = \sum_i \left[ ( 1 + 3 \lambda ) \bsigma_i
+ \frac{ 4 - 3 \lambda }{2} \bfeta_i \right] \cdot \bB~~(B = g_J \mu_{\rm B} H),
\end{align}
where the vectorial octupole operator $\bfeta$ is given by the product of $\btau$ and $\bsigma$
components:
\cite{Shiina97}
\begin{align}
( \eta^x, \eta^y, \eta^z )
= ( - \tau^z \sigma^x + \sqrt{3} \tau^x \sigma^x, - \tau^z \sigma^y - \sqrt{3} \tau^x \sigma^y,
2 \tau^z \sigma^z ).
\end{align}
These components correspond to those in Eq.~(\ref{eqn:G4JT}).
Indeed, the highly symmetric interaction terms can be divided into nonmagnetic and magnetic
sectors as $H_{\rm s} = H_{\rm Q} + H_{\rm M} + H_{\rm Z}$:
\begin{align}
& H_{\rm Q} = D_{\rm Q} \sum_{ \langle i, j \rangle }
[ ( \tilde{\tau}_i^z \tilde{\tau}_j^z + \tilde{\tau}_i^x \tilde{\tau}_j^x )
+ ( \tilde{\tau}_i^y \tilde{\bsigma}_i ) \cdot ( \tilde{\tau}_j^y \tilde{\bsigma}_j ) ]
\nonumber \\
&~~~~~
= \frac{ D_{\rm Q} }{3} \sum_{ \langle i,j \rangle } \sum_\mu O_{i, \mu} O_{j, \mu}~~
(\mu = u, v, yz, zx, xy),
\label{eqn:HQ} \\
& H_{\rm M} = D_{\rm M} \sum_{ \langle i, j \rangle }
[ \tilde{\bsigma}_i \cdot \tilde{\bsigma}_j
+ ( \tilde{\tau}_i^z \tilde{\bsigma}_i ) \cdot ( \tilde{\tau}_j^z \tilde{\bsigma}_j )
+ ( \tilde{\tau}_i^x \tilde{\bsigma}_i ) \cdot ( \tilde{\tau}_j^x \tilde{\bsigma}_j )
+ \tilde{\tau}_i^y \tilde{\tau}_j^y ]
\nonumber \\
&~~~~~
= \frac{4}{5} D_{\rm M} \Bigg\{ \sum_{ \mu = x, y, z } J_{i, \mu} J_{j, \mu}
+ \frac{4}{9} \Bigg[ \sum_{ \mu = x, y, z } ( T_{i, \mu}^\alpha T_{j, \mu}^\alpha
+ T_{i, \mu}^\beta T_{j, \mu}^\beta )
\nonumber \\
&~~~~~~~~~~~~~~~~~~~~~~~~
+ T_{i, xyz} T_{j, xyz} \Bigg] \Bigg\},
\label{eqn:HM}
\end{align}
where $\tilde{\bsigma} = 2 \bsigma$ and $\tilde{\btau} = 2 \btau$.
The coupling constants satisfy $D_{\rm Q} = D_{\rm M} = D$ for the SU(4) symmetric model.

\section{Transition temperatures for the three typical axes of a low magnetic field}
Here, we discuss the octupole effects on the antiferroquadrupole order phase using the GL
expansion of the free energy for the spherical multipole--multipole interaction model in Sect.~3.3.
The anisotropy arising from the octupoles is represented by the parameter $\lambda$ in
Eq.~(\ref{eqn:HZ}).
For a deeper understanding of the octupole effects, it is very useful to compare the GL expansion
for both $\bH \parallel [111]$ and $\bH \parallel [110]$ with that for $\bH \parallel [001]$ studied by
Shiina {\it et al.}.
\cite{Shiina97}

\subsection{$\bH \parallel [001]$ case}
The low-field antiferroquadrupole $O_u$ ($\tau^Z$) phase is described by the relevant staggered
quadrupole moment $\tilde{\tau}_{\rm af}^Z$, other staggered moments
($\tilde{\sigma}_{\rm af}^Z$, $\tilde{\eta}_{\rm af}^Z$), and uniform moments
($\tilde{\tau}_{\rm f}^Z$, $\tilde{\sigma}_{\rm f}^Z$, $\tilde{\eta}_{\rm f}^Z$) induced by the
quadrupole order under the applied magnetic field.
The GL free energy is expanded up to the third order in which three kinds of multipoles
(dipole, quadrupole, and octupole) are coupled to each other.
With the increase in the field strength, the quadrupole order is stabilized by the third order of
the free energy given by
\begin{align}
F_{[001]}^{(3)} = -T ( \tilde{\tau}_{\rm af}^Z \tilde{\sigma}_{\rm f}^Z \tilde{\eta}_{\rm af}^Z
+ \tilde{\eta}_{\rm af}^Z \tilde{\tau}_{\rm f}^Z \tilde{\sigma}_{\rm af}^Z
+ \tilde{\sigma}_{\rm af}^Z \tilde{\eta}_{\rm f}^Z \tilde{\tau}_{\rm af}^Z
+ \tilde{\tau}_{\rm f}^Z \tilde{\sigma}_{\rm f}^Z \tilde{\eta}_{\rm f}^Z ).
\end{align}
Combining $F^{(3)}$ with the second-order and Zeeman terms, we represent the free energy as
\begin{align}
& F - F_0 = \frac{1}{2} ( T - T_0^{\rm Q} ) ( \tilde{\tau}_{\rm af}^Z )^2
+ \frac{1}{2} ( T - T_0^{\rm M} ) \sum_{\gamma = \sigma, \eta} ( \tilde{\gamma}_{\rm af}^Z )^2
\nonumber \\
&~~~~~~~~~~
+ \frac{1}{2} ( T + T_0^{\rm Q} ) ( \tilde{\tau}_{\rm f}^Z )^2
+ \frac{1}{2} ( T + T_0^{\rm M} ) \sum_{\gamma = \sigma, \eta} ( \tilde{\gamma}_{\rm f}^Z )^2
\nonumber \\
&~~~~~~~~~~
+ F_{[001]}^{(3)} - B \left( \frac{ 1 + 3 \lambda }{2} \tilde{\sigma}_{\rm f}^Z
+ \frac{ 4 - 3 \lambda }{4} \tilde{\eta}_{\rm f}^Z \right) - \delta h_{\rm s} \tilde{\tau}_{\rm af}^Z,
\label{eqn:F001}
\end{align}
where $F_0$ is the origin of the free energy for the para-quadrupole phase, and
$T_0^Q \gtrsim T_0^M$ is assumed for the transition temperatures at the zero field.
$\delta h_{\rm s}$ is a fictitious staggered field related to the transition temperature at a finite field
$B$.
\cite{Shiina97}
Minimizing the free energy with respect to $\tilde{\tau}_{\rm af}^Z$, $\tilde{\sigma}_{\rm af}^Z$,
and $\tilde{\eta}_{\rm af}^Z$, we leave the leading terms to determine the transition temperature,
\begin{align}
\left\{ T - T_0^{\rm Q}
- \frac{ T^2 \left[ ( \tilde{\sigma}_{\rm f}^Z )^2 + ( \tilde{\eta}_{\rm f}^Z )^2 \right] }{ T - T_0^{\rm M} }
\right\} \tilde{\tau}_{\rm af}^Z
= \delta h_{\rm s}.
\label{eqn:TQ}
\end{align}
For a finite $B$, the free energy is minimized with respect to $\tilde{\sigma}_{\rm f}$ and
$\tilde{\eta}_{\rm f}$, which leads to the Curie--Weiss law:
\begin{align}
\tilde{\sigma}_{\rm f}^Z = \frac{ 1 + 3 \lambda }{2} \frac{B}{ T + T_0^{\rm M} },~~
\tilde{\eta}_{\rm f}^Z = \frac{ 4 - 3 \lambda }{4} \frac{B}{ T + T_0^{\rm M} }.
\end{align}
The transition temperature at a finite field is given by the solution for $\delta h_s \rightarrow 0$.
After substituting $\tilde{\sigma}_{\rm f}$ and $\tilde{\eta}_{\rm f}$ in Eq.~(\ref{eqn:TQ}), we obtain
the leading contribution of $B$ to the transition temperature:
\begin{align}
T^{\rm Q} = T_0^{\rm Q} + \frac{ 5 ( 4 + 9 \lambda^2 ) }{64} \frac{ B^2 }{ T_0^{\rm Q} - T_0^{\rm M} }.
\end{align}
This indicates that the increase in $T^{\rm Q}$ is proportional to $B^2$ ($B = g_J \mu_{\rm B} H$)
and its proportionality coefficient is increased by the octupole effect with $\lambda$.
In the symmetric limit $T_0^{\rm M} \rightarrow T_0^{\rm Q}$, the transition temperature shows
a linear $B$ dependence
\begin{align}
T^{\rm Q} = T_0^{\rm Q} + \frac{ \sqrt{ 5 ( 4 + 9 \lambda^2 ) } }{8} B.
\end{align}

\subsection{$\bH \parallel [111]$ case}
The above argument for $\bH \parallel [001]$ can be applied to the antiferroquadrupole
$O_{yz} + O_{zx} + O_{xy}$ phase denoted by $\tau^Z$ for $\bH \parallel [111]$ straightforwardly.
The only difference is that the Zeeman term in Eq.~(\ref{eqn:HZ3}) includes the effective
transverse component $( \tilde{\sigma}^X + \tilde{\eta'}^X ) B$ other than $\tilde{\sigma}^Z B$ and
$\tilde{\eta}^Z B$, which leads to the coupling of $| 3/2 \rangle$ and $| - 3/2 \rangle$, as
shown in Table~I.
Accordingly, the additional moments $\tilde{\sigma}^X_{\rm f}$, $\tilde{\eta'}_{\rm f}^X$,
$\tilde{\sigma}^X_{\rm af}$, and $\tilde{\eta'}^X_{\rm af}$ are induced together with
$\tilde{\sigma}^Z_{\rm f}$, $\tilde{\eta}_{\rm f}^Z$, $\tilde{\sigma}^Z_{\rm af}$, and
$\tilde{\eta}^Z_{\rm af}$ simultaneously.
The Zeeman term can be rewritten as
\begin{align}
& H_{{\rm Z}, [111]} = - B \left[ \frac{ 1 - 2 \lambda }{2} \tilde{\sigma}_{\rm f}^Z
+ \frac{ 2 + \lambda }{2} \tilde{\eta}_{\rm f}^Z
+ \frac{ 5 \lambda }{ 4 \sqrt{2} } ( \tilde{\sigma}_{\rm f}^X + \tilde{\eta'}_{\rm f}^X ) \right]
\nonumber \\
&~~~~~~~~~~~~
= - B \left( \frac{G_\lambda}{4} \tilde{\xi}_{\rm f}^+ + \frac{ 1 + 3 \lambda }{2} \tilde{\xi}_{\rm f}^- \right),
\end{align}
where $G_\lambda$ is given in Table~I.
The new parameters $\tilde{\xi}_{\rm f (af)}^\pm$ of uniform (staggered) moments are related to the
above eight moments as
\begin{align}
& \tilde{\sigma}_{\rm f (af)}^Z = \tilde{\xi}_{\rm f (af)}^+ \cos 2 \theta_d - \tilde{\xi}_{\rm f (af)}^-,
\nonumber \\
& \tilde{\eta}_{\rm f (af)}^Z = \tilde{\xi}_{\rm f (af)}^+ \cos 2 \theta_d + \tilde{\xi}_{\rm f (af)}^-,
\nonumber \\
& \tilde{\sigma}_{\rm f (af)}^X = \tilde{\eta'}_{\rm f (af)}^X = \tilde{\xi}_{\rm f (af)}^+ \sin 2 \theta_d,
\nonumber \\
&~~~~~~
(\cos 2 \theta_d = d_+^2 - d_-^2,~~\sin 2 \theta_d = 2 d_+ d_-).
\end{align}
Using $\tilde{\xi}^\pm$ instead of $\tilde{\sigma}^Z$, $\tilde{\eta}^Z$, $\tilde{\sigma}^X$ and
$\tilde{\eta'}^X$, the GL free energy is expanded to the third order,
\begin{align}
& F - F_0 = \frac{1}{2} ( T - T_0^{\rm Q} ) ( \tilde{\tau}_{\rm af}^Z )^2
+ ( T - T_0^{\rm M} ) \sum_{\mu = +,-} ( \tilde{\xi}_{\rm af}^\mu )^2
\nonumber \\
&~~~~~~~~~~~~
+ \frac{1}{2} ( T + T_0^{\rm Q} ) ( \tilde{\tau}_{\rm f}^Z )^2
+ ( T + T_0^{\rm M} ) \sum_{\mu= +,-} ( \tilde{\xi}_{\rm f}^\mu )^2
\nonumber \\
&~~~~~~~~~~~~
+ F_{[111]}^{(3)} + H_{{\rm Z}, [111]} - \delta h_{\rm s} \tilde{\tau}_{\rm af}^Z,
\end{align}
and
\begin{align}
& F_{[111]}^{(3)} = -T ( \tilde{\tau}_{\rm af}^Z \tilde{\sigma}_{\rm f}^Z \tilde{\eta}_{\rm af}^Z
+ \tilde{\tau}_{\rm af}^Z \tilde{\sigma}_{\rm f}^X \tilde{\eta'}_{\rm af}^X
+ \tilde{\eta}_{\rm af}^Z \tilde{\tau}_{\rm f}^Z \tilde{\sigma}_{\rm af}^Z
+ \tilde{\eta'}_{\rm af}^X \tilde{\tau}_{\rm f}^Z \tilde{\sigma}_{\rm af}^X
\nonumber \\
&~~~~~~~~~~~~~~~~
+ \tilde{\sigma}_{\rm af}^Z \tilde{\eta}_{\rm f}^Z \tilde{\tau}_{\rm af}^Z
+ \tilde{\sigma}_{\rm af}^X \tilde{\eta'}_{\rm f}^X \tilde{\tau}_{\rm af}^Z
+ \tilde{\tau}_{\rm f}^Z \tilde{\sigma}_{\rm f}^Z \tilde{\eta}_{\rm f}^Z
+ \tilde{\tau}_{\rm f}^Z \tilde{\sigma}_{\rm f}^X \tilde{\eta'}_{\rm f}^X )
\nonumber \\
&~~~~~~~~~
= - T \Big\{ 2 \tilde{\tau}_{\rm af}^Z ( \tilde{\xi}_{\rm f}^+ \tilde{\xi}_{\rm af}^+
- \tilde{\xi}_{\rm f}^- \tilde{\xi}_{\rm af}^- )
\nonumber \\
&~~~~~~~~~~~~~~~~
+ \tilde{\tau}_{\rm f}^Z \left[ ( \tilde{\xi}_{\rm af}^+ )^2 - ( \tilde{\xi}_{\rm af}^- )^2
+ ( \tilde{\xi}_{\rm f}^+ )^2 - ( \tilde{\xi}_{\rm f}^- )^2 \right] \Big\}.
\end{align}
Minimizing the free energy with respect to $\tilde{\tau}_{\rm af}^Z$, $\tilde{\xi}_{\rm af}^+$, and
$\tilde{\xi}_{\rm af}^-$ at $\delta h_{\rm s} \rightarrow 0$, we obtain the transition temperature
increased by the applied field as
\begin{align}
T^{\rm Q} = T_0^{\rm Q} + \frac{ 2 ( T_0^{\rm Q} )^2 }{ T_0^{\rm Q} - T_0^{\rm M} }
\sum_{\mu = +,-} ( \tilde{\xi}_{\rm f}^\mu )^2.
\end{align}
On the other hand, the Curie--Weiss law is represented by
\begin{align}
\tilde{\xi}_{\rm f}^+ = \frac{G_\lambda}{8} \frac{B}{ T + T_0^{\rm M} },~~
\tilde{\xi}_{\rm f}^- = \frac{ 1 + 3 \lambda }{4} \frac{B}{ T + T_0^{\rm M} }.
\end{align}
Finally, $T^{\rm Q}$ for $T_0^{\rm Q} \gtrsim T_0^{\rm M}$ is obtained as
\begin{align}
T^{\rm Q} = T_0^{\rm Q} + \frac{ 5 ( 4 + 9 \lambda^2 ) }{64} \frac{ B^2 }{ T_0^{\rm Q} - T_0^{\rm M} }.
\end{align}
Note that this is the same as $T^{\rm Q}$ for $\bH \parallel [001]$.
\cite{Shiina97}

\subsection{$\bH \parallel [110]$ case}
For $\bH \parallel [110]$, the Zeeman term includes
$\tilde{\zeta}^Z B = \tilde{\tau}^X \tilde{\sigma}^Z B$, which leads to the coupling of $| 3/2 \rangle$
and $| -1/2 \rangle$ and that of $| 1/2 \rangle$ and $| - 3/2 \rangle$.
As a consequence, the magnetic field can also induce the uniform and staggered quadrupole
moments denoted by $\tau^X$.
Introducing a new parameter $\tilde{\xi}_{\rm f (af)}$ related to the uniform (staggered) moments
$\tilde{\eta}_{\rm f (af)}^Z$ and $\tilde{\zeta}_{\rm f (af)}^Z$ as
\begin{align}
& \tilde{\eta}_{\rm f (af)}^Z = \tilde{\xi}_{\rm f (af)} \cos 2 \theta_c,~~
\tilde{\zeta}_{\rm f (af)}^Z = - \tilde{\xi}_{\rm f (af)} \sin 2 \theta_c,
\nonumber \\
&~~~~~~
( \cos 2 \theta_c = c_+^2 - c_-^2,~~\sin 2 \theta_c = 2 c_+ c_- ),
\end{align}
we can rewrite the Zeeman term as
\begin{align}
& H_{{\rm Z}, [110]} = - \frac{B}{16} \left[ ( 8 - 6 \lambda ) \tilde{\sigma}_{\rm f}^Z
+ ( 16 + 3 \lambda ) \tilde{\eta}_{\rm f}^Z - 15 \sqrt{3} \lambda \tilde{\zeta}_{\rm f}^Z \right]
\nonumber \\
&~~~~~~~~~~~~
= - B \left( \frac{ 4 - 3 \lambda}{8} \tilde{\sigma}_{\rm f}^Z
+ \frac{F_\lambda}{16} \tilde{\xi}_{\rm f} \right),
\end{align}
where $F_\lambda$ is given in Table~I.
In addition, we represent primary order parameters $\tau_{\rm f (af)}^Z$ and $\tau_{\rm f (af)}^X$
by the linear combination of new parameters $\tau'$ and $\tau''$ as
\begin{align}
& \tau_{\rm f (af)}^Z = \tau'_{\rm f (af)} \cos 2 \theta_c - \tau''_{\rm f (af)} \sin 2 \theta_c,
\nonumber \\
& \tau_{\rm f (af)}^X = - \tau'_{\rm f (af)} \sin 2 \theta_c - \tau''_{\rm f (af)} \cos 2 \theta_c.
\end{align}
For the expansion of the GL energy up to the third order, the leading terms are obtained as
\begin{align}
& F - F_0 = \frac{1}{2} ( T - T_0^{\rm Q} ) ( \tilde{\tau'}_{\rm af} )^2
+ \frac{1}{2} ( T - T_0^{\rm M} ) \sum_{\gamma = \sigma^Z,~\xi} \tilde{\gamma}_{\rm af}^2
\nonumber \\
&~~~~~~~~~~~~
+ \frac{1}{2} ( T + T_0^{\rm Q} ) ( \tilde{\tau'}_{\rm f} )^2
+ \frac{1}{2} ( T + T_0^{\rm M} ) \sum_{\gamma = \sigma^Z,~\xi} \tilde{\gamma}_{\rm f}^2
\nonumber \\
&~~~~~~~~~~~~
+ F_{[110]}^{(3)} + H_{{\rm Z}, [110]} - \delta h_{\rm s} \tilde{\tau'}_{\rm af}^Z,
\label{eqn:F110}
\end{align}
and
\begin{align}
& F_{[110]}^{(3)} = -T ( \tilde{\tau}_{\rm af}^Z \tilde{\sigma}_{\rm f}^Z \tilde{\eta}_{\rm af}^Z
+ \tilde{\tau}_{\rm af}^X \tilde{\sigma}_{\rm f}^Z \tilde{\zeta}_{\rm af}^Z
+ \tilde{\eta}_{\rm af}^Z \tilde{\tau}_{\rm f}^Z \tilde{\sigma}_{\rm af}^Z
+ \tilde{\zeta}_{\rm af}^Z \tilde{\tau}_{\rm f}^X \tilde{\sigma}_{\rm af}^Z
\nonumber \\
&~~~~~~~~~~~~~~~~
+ \tilde{\sigma}_{\rm af}^Z \tilde{\eta}_{\rm f}^Z \tilde{\tau}_{\rm af}^Z
+ \tilde{\sigma}_{\rm af}^Z \tilde{\zeta}_{\rm f}^Z \tilde{\tau}_{\rm af}^X
+ \tilde{\tau}_{\rm f}^Z \tilde{\sigma}_{\rm f}^Z \tilde{\eta}_{\rm f}^Z
+ \tilde{\tau}_{\rm f}^X \tilde{\sigma}_{\rm f}^Z \tilde{\zeta}_{\rm f}^Z )
\nonumber \\
&~~~~~~~~~
= -T ( \tilde{\tau'}_{\rm af} \tilde{\sigma}_{\rm f}^Z \tilde{\xi}_{\rm af}
+ \tilde{\xi}_{\rm af} \tilde{\tau'}_{\rm f} \tilde{\sigma}_{\rm af}^Z
+ \tilde{\sigma}_{\rm af}^Z \tilde{\xi}_{\rm f} \tilde{\tau'}_{\rm af}
+ \tilde{\tau'}_{\rm f} \tilde{\sigma}_{\rm f}^Z \tilde{\xi}_{\rm f} ).
\end{align}
We note that the form of Eq.~(\ref{eqn:F110}) is similar to that of Eq.~(\ref{eqn:F001}).
Thus, the transition temperature for $T_0^{\rm Q} \gtrsim T_0^{\rm M}$ is calculated up to the
order of $B^2$ as
\begin{align}
& T^{\rm Q} = T_0^{\rm Q} + \frac{ B^2 }{ 4 ( T_0^{\rm Q} - T_0^{\rm M} ) }
\left[ \left( \frac{ 4 - 3 \lambda }{8} \right)^2 + \left( \frac{F_\lambda}{16} \right)^2 \right]
\nonumber \\
&~~~~~
= T_0^{\rm Q} + \frac{ 5 ( 4 + 9 \lambda^2 ) }{64} \frac{ B^2 }{ T_0^{\rm Q} - T_0^{\rm M} }.
\end{align}
The result is that the increase in $T^{\rm Q}$ for $\bH \parallel [110]$ shows the same field
dependence as that for $\bH \parallel [001]$ and $[111]$.
\cite{Shiina97}
The stabilized quadrupole component is written as
\begin{align}
& \tilde{\tau'} = \tilde{\tau}^Z \cos 2 \theta_c - \tilde{\tau}^X \sin 2 \theta_c
\nonumber \\
&~~~
= \frac{1}{ \sqrt{3} F_\lambda} [ ( 16 + 3 \lambda ) O_U - 15 \sqrt{3} \lambda \cdot O_V ) ],
\end{align}
and it has only diagonal matrix elements for the eigenstates in Table~I:
$\langle \psi_{1 (4)} | \tilde{\tau'} | \psi_{1 (4)} \rangle = 1$
and $\langle \psi_{2 (3)} | \tilde{\tau'} | \psi_{2 (3)} \rangle = - 1$ for $\bH \parallel [110]$.
Using the unitary transformation in Eq.~(\ref{eqn:OkOK}), the two components $O_U$ and $O_V$ of $\tilde{\tau'}$ are represented by the linear combination of
$O_u$ and $O_{xy}$ in the original $(xyz)$ reference frame as
\begin{align}
O_U \rightarrow - \frac{1}{2} O_u + \frac{ \sqrt{3} }{2} O_{xy},~~
O_V \rightarrow \frac{ \sqrt{3} }{2} O_u + \frac{1}{2} O_{xy}
\end{align}
for the SU(4) symmetric interaction model.



\begin{thebibliography}{99}

\bibitem{Santini09}
  P. Santini, S. Carretta, G. Amoretti, R. Caciuffo, N. Magnani, and G. H. Lander,
  Rev. Mod. Phys. {\bf 81}, 807 (2009).
    
\bibitem{Kuramoto09}
  Y. Kuramoto, H. Kusunose, and A. Kiss,
  J. Phys. Soc. Jpn. {\bf 78}, 072001 (2009).
  
\bibitem{Shiina97}
  R. Shiina, H. Shiba, and P. Thalmeier,
  J. Phys. Soc. Jpn. {\bf 66}, 1741 (1997).
     
\bibitem{Hernandez-Minguez21}
  A. Hern\'{a}ndez-M\'{i}nguez, A. V. Poshakinskiy, M. Hollenbach, P. V. Santos, and
  G. V. Astakhov,
  Sci. Adv. {\bf 7}, eabj5030 (2021).
  
\bibitem{Vasselon23}
  T. Vasselon, A. Hern\'{a}ndez-M\'{i}nguez, M. Hollenbach, G. V. Astakhov, and P. V. Santos,
  Phys. Rev. Appl. {\bf 20}, 034017 (2023).
   
\bibitem{Dietz23}
  J. R. Dietz, B. Jiang, A. M. Day, S. A. Bhave, and E. L. Hu,
  Nat. Electron. {\bf 6}, 739 (2023).
  
\bibitem{Hernandez-Minguez20}
  A. Hern\'{a}ndez-M\'{i}nguez, A. V. Poshakinskiy, M. Hollenbach, P. V. Santos, and
  G. V. Astakhov,
  Phys. Rev. Lett. {\bf 125}, 107702 (2020).
  
\bibitem{Koga24}
  M. Koga and M. Matsumoto,
  J. Phys. Soc. Jpn. {\bf 93}, 014703 (2024).
  
\bibitem{Soltamov19}
  V. A. Soltamov, C. Kasper, A. V. Poshakinskiy, A. N. Anisimov, E. N. Mokhov, A. Sperlich,
  S. A. Tarasenko, P. G. Baranov, G. V. Astakhov, and V. Dyakonov,
  Nat. Commun. {\bf 10}, 1678 (2019).
  
\bibitem{Kraus14}
  H. Kraus, V. A. Soltamov, D. Riedel, S. V\"{a}th, F. Fuchs, A. Sperlich, P. G. Baranov, V. Dyakonov,
  and G. V. Astakhov,
  Nat. Phys. {\bf 10}, 157 (2014).
  
\bibitem{Widmann15}
  M. Widmann, S.-Y. Lee, T. Rendler, N. T. Son, H. Fedder, S. Paik, L.-P. Yang, N. Zhao, S. Yang,
  I. Booker, A. Denisenko, M. Jamali, S. A. Momenzadeh, I. Gerhardt, T. Ohshima, A. Gali,
  E. Janz\'{e}n, and J. Wrachtrup,
  Nat. Mater. {\bf 14}, 164 (2015).
  
\bibitem{Simin17}
  D. Simin, H. Kraus, A. Sperlich, T. Ohshima, G.V. Astakhov, and V. Dyakonov,
  Phys. Rev. B {\bf 95}, 161201(R) (2017).
  
\bibitem{Nagy18}
  R. Nagy, M. Widmann, M. Niethammer, D. B. R. Dasari, I. Gerhardt, \"{O}. O. Soykal,
  M. Radulaski, T. Ohshima, J. Vu\v{c}kovi\'{c}, N. T. Son, I. G. Ivanov, S. E. Economou, C. Bonato,
  S.-Y. Lee, and J. Wrachtrup,
  Phys. Rev. Appl. {\bf 9}, 034022 (2018).
  
\bibitem{Castelletto20}
  S. Castelletto and A. Boretti,
  J. Phys.: Photonics {\bf 2}, 022001 (2020).
  
\bibitem{Son20}
  N. T. Son, C. P. Anderson, A. Bourassa, K. C. Miao, C. Babin, M. Widmann, M. Niethammer,
  J. U. Hassan, N. Morioka, I. G. Ivanov, F. Kaiser, J. Wrachtrup, and D. D. Awschalom,
  Appl. Phys. Lett. {\bf 116}, 190501 (2020).
  
\bibitem{Sakai97}
  O. Sakai, R. Shiina, H. Shiba, and P. Thalmeier,
  J. Phys. Soc. Jpn. {\bf 66}, 3005 (1997).
  
\bibitem{Shiina98}
  R. Shiina, O. Sakai, H. Shiba, and P. Thalmeier,
  J. Phys. Soc. Jpn. {\bf 67}, 941 (1998).
  
\bibitem{Shiba99}
  H. Shiba, O. Sakai, and R. Shiina,
  J. Phys. Soc. Jpn. {\bf 68}, 1988 (1999).
  
\bibitem{Mannix05}
  D. Mannix, Y. Tanaka, D. Carbone, N. Bernhoeft, and S. Kunii,
  Phys. Rev. Lett. {\bf 95}, 117206 (2005).
  
\bibitem{Kusunose05}
  H. Kusunose and Y. Kuramoto,
  J. Phys. Soc. Jpn. {\bf 74}, 3139 (2005).
  
\bibitem{Nagao06}
  T. Nagao and J. Igarashi,
  Phys. Rev. B {\bf 74}, 104404 (2006).
  
\bibitem{Matsumura09}
  T. Matsumura, T. Yonemura, K. Kunimori, M. Sera, and F. Iga,
  Phys. Rev. Lett. {\bf 103}, 017203 (2009).
  
\bibitem{Nagao10}
  T. Nagao and J. Igarashi,
  Phys. Rev. B {\bf 82}, 024402 (2010).
  
\bibitem{Matsumura12}
  T. Matsumura, T. Yonemura, K. Kunimori, M. Sera, F. Iga, T. Nagao, and J. Igarashi,
  Phys. Rev. B {\bf 85}, 174417 (2012).
  
\bibitem{Portnichenko20}
  P. Y. Portnichenko, A. Akbari, S. E. Nikitin, A. S. Cameron, A. V. Dukhnenko, V. B. Filipov,
  N. Yu Shitsevalova, P. \v{C}erm\'{a}k, I. Radelytskyi, A. Schneidewind, J. Ollivier, A. Podlesnyak,
  Z. Huesges, J. Xu, A. Ivanov, Y. Sidis, S. Petit, J.-M. Mignot, P. Thalmeier, and D. S. Inosov,
  Phys. Rev. X {\bf 10}, 021010 (2020).
  
\bibitem{Koga20a}
  M. Koga and M. Matsumoto,
  J. Phys. Soc. Jpn. {\bf 89}, 024701 (2020).
  
\bibitem{Matsumoto20}
  M. Matsumoto and M. Koga,
  J. Phys. Soc. Jpn. {\bf 89}, 084702 (2020).
    
\bibitem{Koga20b}
  M. Koga and M. Matsumoto,
  J. Phys. Soc. Jpn. {\bf 89}, 113701 (2020).
  
\bibitem{Koga22}
  M. Koga and M. Matsumoto,
  J. Phys. Soc. Jpn. {\bf 91}, 094709 (2022).
  
\bibitem{Gromov00}
  I. Gromov and A. Schweiger,
  J. Magn. Reson. {\bf 146}, 110 (2000).
  
\bibitem{Kalin04}
  M. K\"{a}lin, I. Gromov, and A. Schweiger,
  Phys. Rev. A {\bf 69}, 033809 (2004).
  
\bibitem{Gyorgy22}
  Z. Gy\"{o}rgy, A. P\'{a}lyi, and G. Sz\'{e}chenyi,
  Phys. Rev. B {\bf 106}, 155412 (2022).
  
\bibitem{Shirley65}
  J. H. Shirley,
  Phys. Rev. {\bf 138}, B979 (1965).
  
\bibitem{Son09}
  S.-K. Son, S. Han, and S. I. Chu,
  Phys. Rev. A {\bf 79}, 032301 (2009).
  
\bibitem{Inui90}
  T. Inui, Y. Tanabe, and Y. Onodera,
  {\it Group Theory and Its Applications in Physics} (Springer, Berlin, 1990).
  
\bibitem{Nakamura94}
  S. Nakamura, T. Goto, S. Kunii, K. Iwashita, and A. Tamaki,
  J. Phys. Soc. Jpn. {\bf 63}, 623 (1994).
  
\bibitem{Takigawa83}
  M. Takigawa, H. Yasuoka, T. Tanaka, and Y. Ishizawa,
  J. Phys. Soc. Jpn. {\bf 52}, 728 (1983).
  
\bibitem{Semeno16}
  A. V. Semeno, M. I. Gilmanov, A. V. Bogach, V. N. Krasnorussky, A. N. Samarin, N. A. Samarin,
  N. E. Sluchanko, N. Yu. Shitsevalova, V. B. Filipov, V. V. Glushkov, and S. V. Demishev,
  Sci. Rep. {\bf 6}, 39196 (2016).
  
\bibitem{Schlottmann18}
  P. Schlottmann,
  Magnetochemistry {\bf 4}, 27 (2018).
  
\bibitem{Semeno21}
  A. V. Semeno, S. Okubo, H. Ohta, and S. V. Demishev,
  Appl. Magn. Reson. {\bf 52}, 459 (2021).
  
\bibitem{Mito23}
  T. Mito, H. Mori, K. Miyamoto, T. Tanaka, Y. Nakai, K. Ueda, F. Iga, and H. Harima,
  J. Phys. Soc. Jpn. {\bf 92}, 034702 (2023).
  
\bibitem{Simin16}
  D. Simin, V. A. Soltamov, A. V. Poshakinskiy, A. N. Anisimov, R. A. Babunts, D. O. Tolmachev,
  E. N. Mokhov, M. Trupke, S. A. Tarasenko, A. Sperlich, P. G. Baranov, V. Dyakonov, and
  G. V. Astakhov,
  Phys. Rev. X {\bf 6}, 031014 (2016).
  
\bibitem{Tarasenko18}
  S. A. Tarasenko, A. V. Poshakinskiy, D. Simin, V. A. Soltamov, E. N. Mokhov, P. G. Baranov,
  V. Dyakonov, and G. V. Astakhov,
  Phys. Status Solidi B {\bf 255}, 1700258 (2018).
  
\bibitem{Udvarhelyi18a}
  P. Udvarhelyi, V. O. Shkolnikov, A. Gali, G. Burkard, and A. P$\acute{\rm a}$lyi,
  Phys. Rev. B {\bf 98}, 075201 (2018).
  
\bibitem{Udvarhelyi18b}
  P. Udvarhelyi and A. Gali,
  Phys. Rev. Appl. {\bf 10}, 054010 (2018).
    
\bibitem{Ohkawa83}
  F. J. Ohkawa,
  J. Phys. Soc. Jpn. {\bf 52}, 3897 (1983).
  
\bibitem{Ohkawa85}
  F. J. Ohkawa,
  J. Phys. Soc. Jpn. {\bf 54}, 3909 (1985).

\end{thebibliography}
\end{document}